\documentclass[aps,twocolumn,eqsecnum,amsmath,amssymb,floatfix,showpacs,
superscriptaddress]{revtex4}
\usepackage{epsfig,psfrag}
\usepackage{amsmath}
\usepackage{amssymb}
\def\avg#1{\langle#1\rangle}

\def\be{\begin{equation}}       \def\ee{\end{equation}}
\def\bea{\begin{eqnarray}}      \def\eea{\end{eqnarray}}
\def\pp{\parallel}

\begin{document}

\title{Competing Orders in Coupled Luttinger Liquids}

\author{Congjun Wu}
\affiliation{Department of Physics, University of Illinois at Urbana-Champaign,
Urbana, Illinois 61801-3080}
\affiliation{Department of Physics, McCullough Building, Stanford University,
 Stanford, California 94305-4045}
\author{W. Vincent Liu}
\affiliation{Department of Physics, University of Illinois at Urbana-Champaign,
 Urbana, Illinois 61801-3080}
\affiliation{Department of Physics, Massachusetts Institute of
Technology,  Cambridge, Massachusetts 02139-4307}
\author{Eduardo Fradkin}
\affiliation{Department of Physics, University of Illinois at Urbana-Champaign,
Urbana, Illinois 61801-3080}
\date{\today}
\begin{abstract}
We consider the problem of two coupled Luttinger liquids both at half filling 
and at low doping levels, to investigate the problem of competing orders in
quasi-one-dimensional strongly correlated systems. 
We use bosonization and renormalization group equations to investigate the
phase diagrams, to determine the allowed phases and to establish approximate
boundaries among them.
Because of the chiral translation and reflection symmetry in the charge mode
away from half filling, orders of charge density wave (CDW) and 
spin-Peierls (SP), diagonal current (DC) and $d$-density wave (DDW) form
two doublets and thus can  be at most quasi-long range ordered.
At half-filling, Umklapp terms break this symmetry down to a discrete group
and thus Ising-type ordered phases appear as a result of spontaneous 
breaking of the residual symmetries. 
Quantum disordered Haldane phases are also found, with finite amplitudes
of pairing orders and triplet counterparts of CDW, SP, DC and DDW.
Relations with  recent numerical results and implications to similar problems
in two dimensions are discussed.
\end{abstract}

\maketitle

\section{Introduction}

The problem of the nature of the phase diagram of the cuprate superconductors
remains at the center of research in the physics of strongly correlated 
electron systems.
Recent work has focused on the possible competing orders responsible for the
known features of the phase diagram as well as to the unusual physical
properties of the pseudo-gap regime.
A number of candidate competing orders have been considered, including 
antiferromagnetism (AF), d-wave pairing (DSC), incommensurate charge ordered
states and other liquid crystal-like phases, and d-density wave states (DDW)
(also known as staggered flux states (SF) or orbital antiferromagnetism (OAF)),
among others.  

$SO(5)$ theory~\cite{zhang} focuses on the competition between 
antiferromagnetism and $d$-wave superconductivity. 
In this theory the natural $SU(2) \times U(1)$  symmetry of the spin and
charge degrees of freedom is regarded as the result of an explicit symmetry
breaking of a larger symmetry, characterized by a global $SO(5)$ group. 
In this picture, this larger symmetry is not apparent except close to a 
quantum critical point whose quantum fluctuations suppress both 
antiferromagnetism and $d$-wave superconductivity, thus leading to a 
pseudo-gap regime controlled by this fixed point.
 
In contrast, in the stripe mechanism~\cite{stripe}, the ground state of the 
doped Mott insulator is an inhomogeneous charge ordered state resembling a 
liquid crystal phase~\cite{nature}, which breaks both rotational invariance 
and (partially) translation invariance, {\it i.\ e.\/} it is a quantum smectic.
In this picture the pseudo-gap is the spin gap which develops in these 
quasi-one-dimensional states, and it is not a signature of some sort of long 
range order. In this mechanism, macroscopic phase coherence and $d$-wave 
superconductivity result from inter-stripe Josephson couplings
\cite{stripe,mats}.  

In the $d$-density wave state, and similarly in the physically equivalent
staggered flux and orbital antiferromagnetic  states, there is a hidden order
which has the same $d_{x^2-y^2}$ symmetry as a $d$-wave superconductor. 
In this phase the ground state has an ordered pattern of staggered orbital
currents, and this is the order which competes with $d$-wave 
superconductivity ~\cite{affleck, wen, ddw-nayak-1, chak1, varma, 
wu}. 

However, in spite of a continued effort during the past decade or so, 
and largely due to the lack of systematic non-perturbative methods in two
dimensions, it has been quite difficult to establish the phase diagram of 
reasonable two dimensional strongly correlated systems based on 
the Hubbard model. Much of the work done is based on mean field-type 
approximations which favor one type of order over others or privileges the
competition among a particular pair of order parameters.
While it is quite possible that these studies reveal different aspects of
possible phase diagrams of some generic, possibly short range models,
it is not possible at present to  determine reliably the phase diagram of many
of these models except sometimes at extreme regimes of  some parameter. 
Thus different approaches, including large-$N$ methods (and their relatives), 
have been used to construct spin-liquid states ~\cite{sr-rvb, pwa, chiral,
rvb-kotliar, affleck-2, ubbens-lee, nodal}. 
Hartree-Fock, large-$d$  and large $N$ methods  have been used to study phase 
separation and striped  states ~\cite{zaanen,schulz-stripe,carlson,sachdev}.
Similarly Hartree-Fock methods have also  been used to study the competition
between superconductivity and DDW order ~\cite{ddw-nayak2}. 
There is also an extensive literature on numerical simulations which work
either at moderate  to high temperatures (as in Quantum Monte Carlo 
simulations due to the fermion sign problem) or exact diagonalizations of 
systems which are usually too small to resolve these issues.    

It is largely for these reasons, as well as for the need of non-perturbative
results, that some of these questions have been considered in the framework 
of quasi-one dimensional systems such as Hubbard-type models (in a loose sense)
on chains and ladders. 
Many of these issues, but not all, can be studied in quasi-one-dimensional 
systems. 
However, not all of these questions can be addressed in one dimension as the
physics may be quite different.
For instance the two-dimensional spin liquid states in two dimensions have 
very specific features with no counterpart in one dimension (not even in
ladders)~\cite{mudry,senthil,sondhi}.
Likewise, the description of a doped one-dimensional Mott insulator at weak 
coupling is a Luttinger liquid while at strong coupling is an incommensurate 
soliton crystal which is also a Luttinger liquid, albeit with strongly 
renormalized parameters. 
In contrast in two dimensions at weak coupling one may expect to find Fermi
liquid pockets while at stronger couplings there is a host of possible liquid
crystal like phases going from a solid to a stripe (or smectic) to a 
nematic whose behavior is markedly different from their one-dimensional
counterparts (when they exist).
Nevertheless, and in spite of these caveats, studies of quasi-one-dimensional 
systems have yielded a wealth information on the physics of strongly
correlated systems.

The simplest quasi-one-dimensional system for the study of some the competing 
orders described above (and others) are ladder systems. 
Away from half filling Hubbard-type models on ladder systems can be reduced to
the problem of two coupled Luttinger liquids. 
There is by now a rather extensive literature on the properties of coupled 
Luttinger liquids. 
These systems have been studied both analytically ~\cite{varma2,fab,rice,
schulz,orignac,shelton,balents,tsvelik,so5-ladder,lin,fjarestad} and 
numerically ~\cite{2leg,2leg-hubbard,2leg-spin-gap,marston,scala,chak2} 
partly for their theoretical simplicity as well as a laboratory to test ideas
intended to work possibly in two dimensions, and for their relevance to
ladder compounds ~\cite{dagotto-rice}. 
As it turns out, systems of two coupled Luttinger liquids can  support almost
all of the local orders proposed for two-dimensional systems and thus shed some
light on them. 
It is thus interesting to investigate this setting the competition between 
different sorts of possible ordered states, to investigate their phase diagrams
systematically and to compare with numerical results.

In this paper we investigate  the phase diagrams of two weakly coupled 
Luttinger liquids both at low doping levels and at half-filling, 
using bosonization and renormalization group (RG) methods.
A number of authors have considered before many aspects of this 
problem (see in particular Refs.\ ~\cite{schulz, orignac, shelton, balents,
tsvelik, lin,fjarestad}). 
Although many of the phases that will discuss here have been discussed before,
we also find a number of new and interesting phases as well as a number of 
new symmetry relations between some of these phases.

One of the motivations of this paper was the recent suggestion that the 
Ising-like order parameter of the ${\mathbb Z}_2$ symmetry of the DDW phase 
could be observed separately from the incommensuration associated with varying 
the doping level~\cite{ddw-nayak-1,chak1,ddw-nayak2}. 
If this was true it may be possible to have a stable phase on a ladder with 
spontaneously broken ${\mathbb Z}_2$. 
Unfortunately, and in agreement with recent results by Fjarestad and Marston
~\cite{fjarestad}, we find that while the DDW order parameter does contain an 
Ising-like piece (as it should) it always involves the charge degree of
freedom which leads to incommensurate behavior. 
On a ladder this leads to correlation function which decays like a power of 
the distance. 
Although our results were derived at weak coupling we expect that this behavior
should extend to strong coupling as well (with the usual large but finite 
renormalizations of velocities and exponents.) 
However, in two dimensions this implies at least two (and possibly more)
possible and distinct phases: a Fermi liquid like DDW phase with pockets
~\cite{ddw-nayak-1,chak1}, and a smectic (or stripe) phase with DDW order. 
We also find that it is quite hard to reach this phase in a ladder system, 
at least within a naive derivation of the effective bosonized theory from 
Hubbard-like microscopic models, which we summarize in the Appendix A. 
Recent, unpublished, numerical simulations by Troyer, Chakravarty and 
Schollw\"ock~\cite{chak2} have reached similar conclusions although in a 
regime where the couplings are larger. 
These authors find exponentially decaying correlations and hence only 
short range DDW order, which means that the simulations reflect a 
quantum disordered phase (of the type described below).
(See also the recent work of Stanescu and Phillips~\cite{tudor}.)

The inter-twinning of charge order with some other sort of order (with a 
discrete symmetry group) is obviously not peculiar to DDW order. 
This is a rather generic situation which leads to interesting phases. 
It also happens for instance, and this is well known, to the Spin-Peierls or 
dimerized phase which upon doping in two dimensions it also becomes either 
a Fermi liquid driven by Fermi surface pockets at weak coupling, or a liquid
crystal phase, such as a stripe state, at intermediate and strong coupling.
One such example is a bond-centered stripe state which was considered at some
length by Vojta, Zhang and Sachdev~\cite{sachdev}, or a site centered stripe 
of the type considered by Granath and co-workers~\cite{mats} which has a rich 
phase diagram. 
In a ladder system these phases are Luttinger liquid which cannot be 
qualitatively distinguished from their weak coupling counterparts.

We also find a number of interesting symmetries relating pairs of these phases.
We find that, away from half filling, the charge density wave phase (CDW) with
the spin-Peirels phase (SP) (or bond-density wave (BW)),  and a new
diagonal current phase (DC)
(described below) with the commensurate DDW phase, form
two doublets under the continuous symmetry of sliding the charge profile, 
represented by the uniform chiral shift of the charge Luttinger field 
$\phi_{c+}$:
$\phi_{c+}\rightarrow \phi_{c+}+\alpha ~ (\mbox{mod}~ 4\sqrt \pi),~
{\phi_{c\pm}\rightarrow -\phi_{c\pm}}$ 
(where the real number $\alpha$ is an arbitrary phase), 
\textit{i.\ e.} a  chiral  translation on a circle and a reflection. 
This continuous symmetry group is non-Abelian and  it may be  denoted by
$C_{\infty v}$, in Schoenflies' symbols. 
Since in one dimensional quantum systems continuous symmetries cannot be 
broken spontaneously, they can only exhibit at most  quasi long range 
fluctuating order and power-law correlations.
However, at half-filling Umklapp terms break the continuous symmetry 
$C_{\infty v}$ down to the finite group $C_{4 v}$, \textit{i.\ e.\/} 
$\phi_{c+}\rightarrow \phi_{c+}+ n\sqrt\pi ~ (\mbox{mod}~ 4\sqrt \pi)$ and
$\phi_{c\pm}\rightarrow -\phi_{c\pm}$.
Hence at half filling these symmetries can be broken spontaneously leading 
to  true long range ordered (LRO) Ising type phases.
In addition we also find four quantum disordered Haldane-like  phases whose low
energy physics can be described by a suitable $O(3)$ non-linear $\sigma$ model.
In these phases there is a spin gap which remains present away from half 
filling. 
In this regime these phases are Luther-Emery liquids. 
There is considerable numerical and analytic evidence for these spin-gap 
phases which in agreement with our conclusions ~\cite{2leg, 2leg-hubbard,
2leg-spin-gap, dagotto-rice}. 
We also discuss in detail the nature of the quantum phase transitions found 
at half-filling.
 
This paper is organized as follows. In Section \ref{sec:model} we present the 
effective Hamiltonians and the order parameters used below to characterize 
the different phases in their bosonized form.
In Section \ref{sec:rg} we use a renormalization group analysis and the known 
strong coupling behaviors of the effective theory at low doping level to
construct a phase diagram.
In Section \ref{sec:half} we do the same type of analysis in Section
\ref{sec:rg} but at half filling. 
In Section \ref{sec:conclusions} we present our conclusions.
In Appendix \ref{sec:ham}
 we relate the parameters of the effective bosonized theory
with those of the extended Hubbard model on the ladder,
and in Appendix \ref{sec:order-p} we give explicit expressions
for the order parameters of interest in terms of the bosonic fields.

\section{Model Hamiltonians and Order Parameters }
\label{sec:model}
We begin with two coupled one-dimensional chains.
To a large extent we will follow the approach used  by Schulz in Ref.\ 
\cite{schulz}.
We consider first the non-interacting limit, and diagonalize the kinetic part
in terms of  ``bonding" and ``anti-bonding" bands (denoted by $1$ and $2$ 
respectively), \textit{i.\ e.\/} symmetric and antisymmetric  under the exchange
of the chain labels.
Including nearest neighbor (NN) hopping, the non-interacting dispersion 
relations are just $\epsilon_{i\sigma}(k)=-2t\cos k\mp t_\perp (i=1,2)$,
where  $t_\perp$ is  inter-chain hopping integral.
This approach makes sense if $t_\perp$ is large compared to any of the 
dynamically generated gaps of the system, \textit{i.\ e.\/} in the weakly 
interacting limit. 

To first order in the doping level $\delta$, the
Fermi wavevectors of two bands are, respectively, 
$k_{f1,2} a =\pi{(1-\delta)/ 2}\pm\sin^{-1}(t_\perp / 2 t)$,
and the corresponding bare Fermi velocities are 
$v_{f1,2}/a= \sqrt{ 4t^2-t_\perp^2 }\pm t_\perp \delta \pi/2$,
where $a$ is the lattice constant which will  serve as the short 
distance cutoff in the bosonized theory.
We will consider the regimes of both low doping and half-filling 
(discussed in Section \ref{sec:rg} and Section \ref{sec:half} respectively)
and  we will assume that $t_\perp$ is not necessarily small.
At half filling where the Umklapp processes dominate,
the system has the particle-hole symmetry 
\be
v_{f1}= v_{f2} ~~\mbox{and} ~~k_{f1}+k_{f2}= \pi. \label{ph}
\ee
Away from half-filling, we will assume that the doping level $\delta$ is 
large enough to suppress the effects of all Umklapp processes
(See section \ref{sec:rg}).
However, if $\delta$ is  relatively small, 
the relation Eq. \ref{ph} still holds approximately.
In this regime the difference in their Fermi velocities does not play a very 
important role (see however the discussion in Ref.\ \cite{multi}). 
However as the filling factor of one of the bands approaches zero, the
respective Fermi velocity becomes very small and the physics is 
somewhat changed. 
In this limit there is an enhancement of the processes leading to the formation
of a spin gap~\cite{stripe,balents}. 
Since we will also find spin gap phases we will ignore here this
special regime since it leads to the same physics (albeit with very different 
parameters). 

The effective theory consists then of two coupled Luttinger liquids, for the
bonding and anti-bonding bands, and a set of perturbations, which we describe 
below, each associated with a particular coupling constant.  
In Appendix \ref{sec:ham}, we will relate these coupling constants with the
interaction parameters of an extended Hubbard model on a ladder with
hopping amplitudes $t$ and $t_\perp$, on-site Hubbard repulsion $U$, and 
Coulomb interactions $V_\parallel$ (on the chains), $V_\perp$ (on the rungs) 
and $V_d$ (along the diagonals of the elementary plaquette), 
as well as  the  exchange Heisenberg interactions $J_\parallel$
(on the chains) and  $J_\perp$ (on the rungs). 

We bosonize the effective theory by introducing a charge bose field and a spin
bose field for both the bonding and anti-bonding Fermi fields, $\phi_{\nu,i}$,
where $i=1,2$ and $\nu=c,s$,  
where $c$ and $s$ label charge and spin modes respectively. 
These fields are mixed under the effects of various interactions, in particular
the backscattering coupling of the respective charge and spin currents and
densities. 
The bosonized theory is diagonalized in terms of the even and odd combinations
of  bose fields from each band: 
$\phi_{\nu\pm}=(\phi_{\nu,1}\pm\phi_{\nu,2})/\sqrt{2}$,
$\theta_{\nu\pm}=(\theta_{\nu,1}\pm\theta_{\nu,2})/\sqrt{2}$, {$\nu=c,s$}.

The quadratic parts of the Hamiltonian density has the standard ``universal" 
form:
\begin{eqnarray}
{\cal H}_{c,\pm}&=& \frac{v_{c,\pm}}{2}\left[
K_{c,\pm}
\Pi_{c,\pm}^2+
\frac{1}{K_{c,\pm}}
(\partial_x \phi_{c,\pm})^2
\right]
\nonumber\\
{\cal H}_{s\pm}&=& 
\frac{v_{s,\pm}}{2}\left[
K_{s,\pm}
\Pi_{s,\pm}^2+
\frac {1}{K_{s,\pm}}
(\partial_x \phi_{s,\pm})^2
\right] \hspace{5mm}
\label{Hfree}
\end{eqnarray}
where $\Pi_{\nu,\pm}$ are the momenta canonically conjugate to the bose fields 
$\phi_{\nu,\pm}$. 
The effective Luttinger parameters and velocities $v_{c,\pm}$ and $v_{s,\pm}$ 
are given by
\begin{eqnarray}
&&K_{c\pm}=\sqrt{ \frac{2 \pi v_f\mp {g_{c\pm}} }
{2 \pi v_f\pm {g_{c\pm}}} },
~~
K_{s\pm}=\sqrt{ \frac{2 \pi v_f\pm {g_{s\pm}} }
{ 2 \pi v_f\mp {g_{s\pm}}} }
\nonumber \\
&&
v_{c,\pm}=\sqrt{ v_f^2-\big ({ g_{c\pm}\over 2\pi} \big)^2},
~~v_{s,\pm}=\sqrt{ v_f^2-\big ({ g_{s\pm}\over 2\pi} \big)^2},
\nonumber\\
\label{eq:Ks}
\end{eqnarray}
where $v_f=(v_{f1}+v_{f2})/2$. The coupling constants $g_{c\pm}$, $g_{s\pm}$ 
correspond to forward-scattering non-chiral couplings of the charge
and spin currents, and are already taken into account in the quadratic terms.
Here we have ignored the effects of straightforward effects of 
forward-scattering chiral couplings, since they only renormalize Fermi 
velocities, and modify the naively determined values of the Luttinger 
parameters. 
Also notice that these expressions can only be taken seriously at weak 
coupling. At intermediate and strong couplings there are also finite but 
significant renormalization of both the Luttinger parameters and of the
velocities.

Let us now discuss the  non-quadratic, interaction terms.
Throughout we will use Majorana Klein factors obeying the convention 
$\eta_\uparrow(1)\eta_\downarrow(1)
\eta_\downarrow(2)\eta_\uparrow(2)=1$.
The backscattering and pair tunneling terms yield the bosonized expressions
\begin{eqnarray}
{\cal H}_{int}&=& \frac{\cos \sqrt{4\pi} \phi_{s+} }{ 2(\pi a)^2 }
(g_1 \cos \sqrt {4\pi} \phi_{s-}-g_2 \cos \sqrt{4\pi} \theta_{s-})\nonumber\\
&+& \frac{ \cos \sqrt{4\pi} \theta_{c-}  }{ 2(\pi a)^2 }
(g_3  \cos \sqrt {4\pi} \theta_{s-}
+g_4  \cos \sqrt{4\pi} \phi_{s-} \nonumber\\
&&\hspace{18mm}+  g_5  \cos \sqrt{4\pi} \phi_{s+}),  
\label{Hint}
\end{eqnarray}
where $\theta_{c,\pm}$ and $\theta_{s,\pm}$
are the dual fields of the charge bosons $\phi_{c,\pm}$ and spin bosons
$\phi_{s,\pm}$ respectively.
Terms labeled by the effective coupling constants $g_1$ and $g_2$ originate 
from the intra-band and inter-band back-scattering interactions $
-g_1 (J^{x,y}_{1R} J^{x,y}_{1L}+{1\rightarrow2})$ and 
$-g_2(J^{x,y}_{1R} J^{x,y}_{2L}+{1\leftrightarrow2})$ respectively.
The terms labeled by the couplings $g_3$, $g_4$ and $g_5$  represent singlet 
and triplet pair-tunneling processes
$\lambda_s (\Delta_1^\dagger \Delta_2 +h.c.)$ and
$\lambda_t (\vec{\Delta}_1^\dagger \vec{\Delta}_2 +h.\ c.\ )$
with $g_3=2 \lambda_t$, $ g_{4}=\lambda_s + \lambda_t$ and $ g_{5}=
\lambda_s - \lambda_t$.
Three  conditions, required by the $SU(2)$ spin rotation invariance, relate 
the spin current and triplet tunneling couplings: $g_{s\pm}=(g_1\pm g_2)/2$
(see also Eq.\  (\ref{eq:Ks})) and $g_5=g_4-g_3$.

\begin{figure}
\centering \epsfig{file=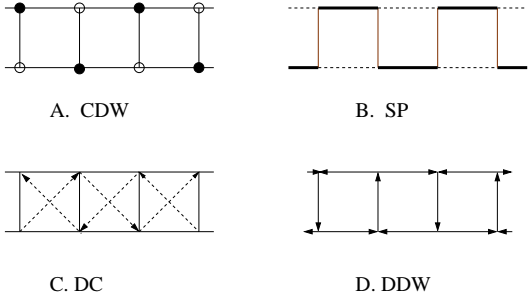,clip=1,width=7cm,angle=0}
\caption{Four Ising type phases.
A. charge density wave(CDW), B. spin-Peierls (SP);
C. diagonal current (DC), D. d-density wave (DDW).
Their triplet analogs are denoted as SDW, SP$^t$;  DC$^t$, DDW$^t$
respectively.
}
\label{order}
\end{figure}

Near half-filling, the following additional Umklapp terms appear as
\begin{eqnarray}
{\cal H}_{um}&=& \frac{ \cos (\sqrt{4\pi} \phi_{c+}-2\delta \pi x)}
{ 2(\pi a)^2 } 
(g_{uc}  \cos \sqrt{4\pi} \theta_{c-} \nonumber \\
&-&g_{u3} \cos \sqrt {4\pi} \theta_{s-}
-g_{u4} \cos \sqrt {4\pi} \phi_{s-} \nonumber \\
&-&g_{u5}  \cos \sqrt {4\pi} \phi_{s+} )  
\label{HUmklapp}
\end{eqnarray}
The term with coupling constant $g_{uc}$ is the so-called ``$\eta$ pair'' 
tunneling processes, \textit{i.\ e.\/} tunneling of Cooper pairs with momentum
$2k_f$, which has the form $m_{R1}^\dagger m_{L2}+(1\rightarrow 2)+h.c.$, 
where $m_{R,L}=\psi_{R,L\uparrow}\psi_{R,L\downarrow}$.
The terms with coupling constants  $g_{u3}$, $ g_{u4}$ and $ g_{u5}$ represent
the couplings between the respective CDW and spin density wave
(SDW) couplings on each chain: $\lambda_{cdw} (N^\dagger(1) N^\dagger(2)
+h.c.)$, $\lambda_{sdw} ({\vec N}^\dagger(1) {\vec N}^\dagger(2) +h.c.)$, 
where $N(i)$ is the $2k_F$ CDW order parameter of chain $i=1,2$, and 
$\vec N(i)$ is the $2k_F$ (N\'eel) SDW order parameter of chain $i=1,2$. 
The coupling constants are $g_{u3}=-\lambda_{sdw}, g_{u4,5}= 
(2\lambda_{cdw}\mp\lambda_{sdw}/2)$.
Due to the $SU(2)$ spin symmetry the condition  $g_{u5}=g_{u4}-g_{u3}$ 
also holds.

For the two-leg  ladder, we  only consider  where  repulsive interactions 
dominate, which implies that the bare values of the effective Luttinger
parameters are in the regime $K_{c+}(0)\ll 1$, $K_{c-}(0), K_{s-}(0)\sim 1$.
Compared with Ref.\ \cite{schulz}, $K_{c-}(0), K_{s-}(0)$  are not necessarily
1, for here they are determined by off-site interactions (see Appendix
\ref{sec:ham}).

Bosonic expressions for various order parameters are given in Appendix
\ref{sec:order-p}.
In the particle-hole (p-h) channel, the possible singlet fermionic bilinear
forms which break the translational symmetry are the order parameters for 
the CDW and SP, DC and DDW operators as shown in Fig.\ \ref{order}. 
The CDW and SP order parameters  are proportional to the real and imaginary 
parts of the symmetric bilinear $\psi^\dagger_{1L\sigma}\psi_{2R\sigma}+
\psi^\dagger_{2L\sigma}\psi_{1R\sigma}$, whereas the DC and DDW order
parameters are the real and imaginary parts of the anti-symmetric version of 
this bilinear. 

From their bosonic representations, we find that all four order parameters 
transform non-trivially under the symmetries broken in their associated phases
(or ground states). 
Thus, for instance the SP and DDW order parameters are odd under the 
${\mathbb Z}_2$ symmetries broken spontaneously by the SP and DDW phases. 
However, in all four cases, these order parameters also involve a phase 
factor (or vertex operator) of the charge boson $\phi_{c,+}$. 
Hence these order parameters also transform non-trivially under shifts of the
charge boson $\phi_{c,+}$, \textit{i.\ e.\/} uniform displacements of the 
charge profile. 
This dependence means that the discrete symmetries, broken spontaneously in
these phases with long range order, are inter-twinned with the continuous 
symmetry of the incommensurate doped state. 
Consequently, these order parameters do not truly acquire an expectation value
but instead only display power law correlations.
Also, while it is possible to write down bosonic expressions for operators 
which transform only under the discrete symmetries broken by these phases,
their fermionic versions are strongly non-local. 
Hence, we conclude that these orders are always incommensurate.

We also find that these order parameters also form two doublets of the 
$C_{\infty v}$ group.
Similarly, their triplet counterparts SDW, SP$^t$, DC$^t$, DDW$^t$
are proportional to real and imaginary parts of
$\psi^\dagger_{1L\alpha}(\vec {\frac{\sigma}{2}})_{\alpha\beta} 
\psi_{2R\beta}\pm\psi^\dagger_{2L\alpha} (\vec {\frac{\sigma}{2}})_{\alpha
\beta}  \psi_{1R\beta}$ respectively (where the label $t$ means triplet).
In the particle-particle (p-p) channel, the  s and d-wave  pairing order
parameters are $\Delta_{s,d}= \sum_{\sigma} (-)^\sigma (\psi_{1L\sigma}
\psi_{1R \bar\sigma} \pm \psi_{2L\sigma}\psi_{2R\bar\sigma})$.
In the next section we identify the stable fixed points of the renormalization
group (RG) flows for the these phases associated with these order parameters.

Some of the order parameters discussed above have been investigated before in
Ref. \cite{schulz, orignac} although under different names. For example,
our $CDW$, $DDW$, $SDW$,  $SSC$ and $DSC$ order parameters are called
$CDW^\pi$, $OAF$, $SDW^\pi$, $SC^s$ and $SC^d$ there.
We note that in a recent paper Ref. \cite{tsuchiizu},
the phases that we label as DDW, SP and DC are called
called $SF$, P-density wave ($PDW$), F-density wave($FDW$) respectively.

Finally, in Eq. \ref{ph}, we ignored the effects of the following terms
\bea
&& \hspace{-8mm}\Delta {\cal H}_c=(\Delta v_f +{\Delta g_c \over \pi}) 
\partial_x \phi_{c+} \partial_x \phi_{c-}
+(\Delta v_f-{\Delta g_c \over \pi}) \Pi_{c+} \Pi_{c-},
\label{broken-ph1}
\nonumber \\ \\
&&  \hspace{-8mm}\Delta {\cal H}_s= (\Delta v_f-{\Delta g_s \over \pi})
\partial_x \phi_{s+} \partial_x \phi_{s-}+(\Delta v_f+
{\Delta g_s \over \pi}) \Pi_{s+} \Pi_{s-}, \nonumber \\
&-& {\Delta g_s \over 2(\pi a)^2} \sin \sqrt {4\pi} \phi_{s-} 
\sin \sqrt {4\pi} \phi_{s+}
\label{broken-ph2}
\nonumber \\ \\
&& \hspace{-8mm} \Delta{\cal H}_{um}= \frac{ \sin (\sqrt{4\pi} \phi_{c+}
-2\delta \pi x)} { 2(\pi a)^2 } (\Delta g_{uc}  \cos \sqrt{4\pi} 
\theta_{c-} \nonumber \\
&-&\Delta g_{u3} \cos \sqrt {4\pi} \theta_{s-}
-\Delta g_{u4} \cos \sqrt {4\pi} \phi_{s-} \nonumber \\
&-& \Delta g_{u5}  \cos \sqrt {4\pi} \phi_{s+} ),  
\label{broken-ph3}
\eea
where $\Delta v_f= \delta \pi t_\perp  /2$ and all other residue
coupling constants varnish linearly with doping near half-filling
as given in Appendix A.
The quadratic residual terms in  Eq. \ref{broken-ph1} and Eq. \ref{broken-ph2}
are marginal perturbations, and they slightly change the scaling 
dimensions of various operators in Eq. \ref{Hint}, \ref{HUmklapp}.
Because they are small, we do not expect that they can change
the stable RG fixed points associated with various phases qualitatively.
For the term of $\Delta g_s$ in Eq. \ref{broken-ph2},  
$\theta_{s+}$ is fixed around $0$ or $\sqrt\pi/2$ at all the stable fixed 
points (see Table \ref{stable-quasi} and \ref{table:Umklapp-stable} below).
The residual Umklapp terms in Eq. \ref{broken-ph3} 
are irrelevant away from half-filling.
At half-filling, $\theta_{c+}$ is fixed at $\sqrt\pi/2$ 
(see Table \ref{table:Umklapp-stable}).
Thus we conclude that all the non-quadratic operators
are irrelevant at all the stable fixed points.
Balents and Fisher \cite{balents} used a perturbative RG of the fermionic
theory and found that a spin gap phase develops near half-filling,
which is consistent with the argument given above.
On the other hand, the continuous $C_{v\infty}$ symmetry is
preserved away from half-filling where the Umklapp terms are irrelevant.
Thus the conclusion that the CDW and SP, DDW and DC order parameters
are incommensurate and thus exhibit quasi long range order is not affected by 
these terms.
However, these residual terms do affect the boundaries among phases. 

\section{Phase Diagram in the Incommensurate Regime}
\label{sec:rg}
We will now investigate the phase diagram in the incommensurate regime, 
but only at low doping. 
In this regime, the Umklapp processes are cut off at a high energy scale
of $2\pi v_f\delta/a$, and can only yield renormalization of the parameters,
such as the velocities, coupling constants and Luttinger parameters
of the low energy effective theory.
The contributions from the Umklapp terms in the RG equations away 
from half-filling \cite{giamarchi} are given in terms of Bessel functions,
which oscillate when a energy scale lower than that of the Umklapp process
is reached.
At this scale, the effects of these terms can be neglected.
Below, we  begin directly at  the low energy scale with all the
coupling constants and Luttinger parameters already renormalized by
the Umklapp terms.

We will investigate the role of the remaining interactions by means of a 
one-loop renormalization group (RG) analysis combined with semiclassical 
arguments. In this regime, the charge boson $\phi_{c,+}$ essentially decouples
and remains gapless. Thus, to one loop order, the Luttinger parameter $K_{c,+}$
does not flow. (This argument is not completely correct: there are always 
irrelevant couplings which do lead to finite renormalizations of $K_{c,+}$;
these effects do not show up at one-loop order.)

The one-loop RG equations for the coupling constants $g_1$ through $g_5$ 
and Luttinger parameters 
 $K_{c,-}$ and $K_{s,\pm}$ are
\begin{eqnarray}
&&\frac{d K_{c-} }{ d l}=\frac{1 }{ 8 \pi^2} (g_3^2 +g_4^2+g_5^2)
\nonumber \\
&&\frac{d K_{s+} }{ d l}=-\frac{K_{s+}^2 }{ 8 \pi^2} (g_1^2 +g_2^2+g_5^2)
\nonumber\\
&&\frac{d K_{s-}}{ d l}=-\frac{K_{s-}^2 }{ 8 \pi^2} (g_1^2 +g_4^2)+
\frac{1}{ 8 \pi^2} (g_2^2+g_3^2) \nonumber \\
&&\frac{d g_1 }{ d l} = (2-K_{s+}-K_{s-}) g_1-\frac{g_4 g_5 }{ 2\pi}
\nonumber \\
&&\frac{d g_2 }{ d l} = (2-K_{s+}-\frac{1}{K_{s-}}) g_2+\frac{g_3 g_5 }{ 2 \pi}
\nonumber \\
&&\frac{d g_3 }{ d l} = (2-\frac{1}{K_{c-}}-\frac{1}{K_{s-}}) g_3 +
\frac{g_2 g_5 }{ 2 \pi} \nonumber \\ 
&&\frac{d g_4 }{ d l}= (2-\frac{1}{K_{c-}}-K_{s-}) g_4
-\frac{g_1 g_5 }{ 2 \pi} \nonumber \\
&&\frac{d g_5 }{ d l}= (2-\frac{1}{K_{c-}}-K_{s+}) g_5
-\frac{g_1 g_4 }{ 2\pi} +\frac{g_2 g_3 }{ 2 \pi}, \nonumber\\
&&
\label{rg-i}
\end{eqnarray}
where $l=\ln(L/a)$ with  the length scale $L$. 

Along the $SU(2)$-invariant manifold for the spin current and pair tunneling 
terms, the RG equations  can be simplified to
\begin{eqnarray}\label{rg1}
&&\frac{d K_{c-} }{ d l}=\frac{1 }{ 8 \pi^2} (g_3^2 +g_4^2+(g_3-g_4)^2),
\nonumber\\
&&\frac{ d g_{s+} }{ d l }=-\frac{1}{ 2 \pi} (g_{s+}^2+g_{s-}^2)-
\frac{(g_3-g_4)^2 }{ 4 \pi},
\nonumber \\
&&\frac{d g_{s-} }{ d l}= -\frac{1}{ \pi} g_{s+} g_{s-} +\frac{g_3^2 }{ 4 \pi}
-\frac{g_4^2}{ 4 \pi},\nonumber \\
&&\hspace{-3mm}
\frac{d g_3 }{ d l}=( 1-\frac{1}{ K_{c-}}+ \frac{-g_{s+} +2g_{s-} }{ 2 \pi} )
g_3+\frac{(g_{s+} -g_{s-}) g_4 }{ 2 \pi}, \nonumber \\
&& \hspace{-3mm}
\frac{d g_4 }{ d l}= ( 1-\frac{1}{ K_{c-}}+ \frac{-g_{s+} -2g_{s-} }{ 2 \pi} )
g_4+\frac{(g_{s+} +g_{s-}) g_3 }{ 2\pi},\nonumber \\
&&
\label{rg-i-su2}
\end{eqnarray}
with 
\begin{equation}
\frac{d}{dl}(g_3-g_4+g_5) = ( 1-\frac{1}{K_{c-}}+\frac{g_{s+}}{2\pi} )
(g_3-g_4+g_5) 
\equiv 0 
\end{equation}
These equations are invariant under transformations
$(g_1,g_2,g_3,g_4)\rightarrow (g_1,g_2,-g_3,-g_4)
\rightarrow (g_2,g_1,g_4,g_3)$.
This means phase boundaries must also have such symmetries.

\begin{table}[h]\label{stbpt}
\begin{center}
\begin{tabular}{|c|c|c|c|c|c|c|c|}   \hline
 & $g_1,g_2$   & $g_3, g_4, g_5$        
 & $\phi_{s+}$ & $\phi_{s-}$ &$\theta_{s-}$ & Order&dimension \\  \hline
1& $0, -\infty$ &  $+\infty, 0, -\infty$ & 0 &/ & 
$\frac{\sqrt \pi }{ 2}$ & CDW+SP&
$ K_{c+}/4$\\ \hline
2& $0, -\infty$ &  $-\infty, 0, +\infty$ 
& $\frac{\sqrt \pi }{ 2}$ &/& 0& DC+DDW&
$ K_{c+}/4$ \\ \hline
3& $-\infty,0$ & $0, +\infty,+\infty$ &
$\frac{\sqrt \pi }{ 2}$  & $\frac{\sqrt \pi }{ 2}  $&/& DSC&
$1/(4 K_{c+})$ \\ \hline
4& $-\infty,0$ & $0, -\infty,-\infty$ & 0&0&/& SSC&$1/(4 K_{c+})$ \\ \hline
\end{tabular} 
\caption{
Stable fixed points and corresponding quasi long range orders away from
half-filling, with $\langle \theta_{c-} \rangle=0$ and
$g_5=g_4-g_3$ (required by $SU(2)$ invariance.)}
\label{stable-quasi}
\end{center}

\begin{center}
\begin{tabular}{|c|c|c|c|c|}   \hline
 & $g_{s+},g_{s-}$& $g_3, g_4, g_5$ & $\phi_{s+}$ & Transition \\  \hline
1&  0,0 & $+\infty, +\infty, 0$ & unfixed &
CDW+SP $\leftrightarrow$ DSC \\ \hline
2&  0,0 & $-\infty,-\infty,0$ & unfixed&
DDW+DC $\leftrightarrow$ SSC \\ \hline
3& $-\infty$,  0& $-\infty,+\infty,+\infty$&$\sqrt \pi/2$&  
DDW+DC $\leftrightarrow$ DSC \\ \hline
4& $-\infty$,  0& $+\infty,-\infty,-\infty$&$0$&  
CDW+SP  $\leftrightarrow$ DSC \\ \hline
\end{tabular} 
\caption{
Critical phase boundaries and unstable fixed points away from half-filling,  
also with $\langle \theta_{c-} \rangle=0$ and
$g_5=g_4-g_3$.}
\label{boundary-quasi}
\end{center}
\end{table}

For ``bare values" of the Luttinger parameter $K_{c-}(0)\sim 1$, the 
marginally relevant RG flow of Eq.\  (\ref{rg-i-su2}) is such that a gap 
develops in the $c-$ sector, which scales like $m_{c-}\approx \exp(-1/g(0))$,
where $g$ is the most relevant one among the marginally relevant perturbations 
$g_{3}$, $ g_{4}$ and $g_{5}$. In this regime $K_{c,-}$ flows to large 
values and, thus from now on we will set $1/K_{c-}=0$. 
In this phase the operator $\cos(\sqrt{4\pi}\theta_{c,-})$ acquires a
non-vanishing expectation value which classically are just $\pm 1$. 
Hence in this phase the dual field takes the values $\theta_{c,-}=0,
\sqrt{\pi}/2$, which are related to each other by a ${\mathbb Z}_2$ 
symmetry~\cite{fjarestad}.
In what follows in this section,  we will choose the 
value $\avg{\theta_{c-}}=0$.

From now on we will use the set  $(g_1,g_2,g_3,g_4)$ to represent the 
stable fixed points of Eq.\ (\ref{rg1}), which are summarized in 
Table ~\ref{stable-quasi}.
At the fixed points $(0, -\infty, \mp\infty,0)$, 
the inter-band  back-scattering coupling constant $g_2$ is relevant, 
while the intra-band back-scattering coupling constant $g_1$  is irrelevant.
Both $\lambda_s$ and $\lambda_t$  are relevant and satisfy the relation
$\lambda_s=-\lambda_t$. 
By direct inspection of their scaling dimensions, we find that
$\lambda_s$ and $\lambda_t$ are more relevant than $g_2$. 
The resulting phase depends on where the RG flows go.
When $g_3\rightarrow -\infty $, the expectation value of $\phi_{s,+}$ and 
$\theta_{s,-}$ asymptotically take the values $\langle  \phi_{s+}\rangle=
\sqrt\pi/2 $ and $\langle  \theta_{s-}\rangle=0$. 
This is the stable fixed point for either the DDW phase or the DC phase.
However, this is true only quasi-long range order (QLRO) due to the strong
fluctuations of the gapless charge boson $\phi_{c,+}$. 
In this phase these order parameters have scaling dimension $K_{c,+}/4$.
Conversely, when $g_3\rightarrow + \infty $, $\langle  \phi_{s+}\rangle=0 $, 
and $\langle  \theta_{s-}\rangle=\sqrt\pi/2$. 
Hence, at this fixed point we would have (naively) either a CDW phase or 
a spin-Peierls (SP (or dimerized) phase.
Here too there is only QLRO and the associated order parameters also have
scaling dimension $K_{c,+}/4$.

We  conclude, in agreement with  the recent results of
Ref.\  \cite{fjarestad},  that the chiral translation symmetry in the field
$\phi_{c,+}$, in other terms due to the charge incommensurability, there 
is no true long range order of the DDW order but only (incommensurate)
power law correlations.
We further can see that the DDW and DC phases ( and also the CDW and SP phases)
form doublet representation under the $C_{\infty v}$ group and are thus 
degenerate. 
Equivalently, the DDW and DC order parameters can be regarded as the real 
and imaginary parts of a single complex order parameter which can thus be
rotated continuously into each other. 
The same relationship holds for the CDW and spin-Peierls order parameters. 
Thus both stable phases, CDW+SP and DDW+DC, have a continuous $U(1)$ symmetry. 
Naturally, since the ladder is a one-dimensional system, this symmetry is no
truly spontaneously broken as there are only power-law correlations for 
these order parameters. 
However, we will see in Section \ref{sec:half} that at half filling,
the Umklapp terms break this symmetry explicitly from $U(1)$ down to
${\mathbb Z}_2$ leading to additional Ising-like phase transitions.

Similarly,  we also find that $\lambda_s$ is more relevant than $g_1$  at
$(-\infty,0,0,\pm\infty)$ while  $g_2$ and $\lambda_t$ are irrelevant.
When $g_4\rightarrow +\infty$,~ $\langle  \phi_{s+}\rangle$ and 
$\langle   \phi_{s-} \rangle$ are fixed at $ \sqrt \pi/2$. 
Thus, d-wave superconductivity (DSC) is the leading QLRO and its order 
parameter has  scaling dimension $1/(4 K_{c+})$. 
Conversely, when $g_4\rightarrow -\infty$,~ $\langle  \phi_{s+}\rangle$ and 
$\langle   \phi_{s-} \rangle$ are fixed at $0$,  s-wave superconductivity
(SSC) is the leading QLRO and its order parameter also has scaling dimension
$1/(4 K_{c+})$. 

\begin{figure} 
\centering \epsfig{file=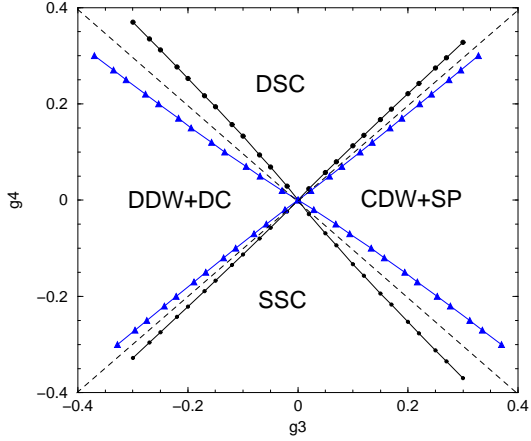,clip=,width=7cm,angle=0}
\caption{
Phase boundaries with positive initial value of $g_{s+}$ ( $g_{s+}(0)=0.2$)
and  different initial values of $g_{s-}(0)$ with dashed line
($g_{s_-}(0)=0$), solid points ($g_{s_-}(0)=0.1$) and triangles 
($g_{s_-}(0)=-0.1$).
Phase boundaries of CDW+SP $\leftrightarrow$ DSC, DDW+DC $\leftrightarrow$
SSC become of first order for $g_{s+}(0)<0$.
}
\label{phase}
\end{figure}  

Let us consider now the phase boundaries and the nature of the phase 
transitions between these possible states, at $g_{s-}(0)=0$. 
In this regime it is more natural to represent instead the unstable fixed 
points with $(g_{s+},g_{s-},g_3,g_4)$.
The RG flows starting with $g_{s+}(0)>0, g_{s-}(0)=0$ and $g_3(0)=g_4(0)=g>0$
evolve towards the fixed point at $(0,0,+\infty,+\infty)$.
Here, the field $\phi_{s+}$ becomes free, $K_{s\pm} \to 1$, and the 
residual interactions reduce to
\begin{eqnarray}\label{eff1}
{\cal H}^1_{res}\!\!\!\!\!\!&&= \frac{g^*}{ 2(\pi a)^2} \avg{\cos
\sqrt {4\pi} \theta_{c-} }
(\cos \sqrt {4\pi} \theta_{s-}+\cos \sqrt {4\pi} \phi_{s-}),
\nonumber \\
&&
\end{eqnarray}
where $g^*$ means the renormalized value of $g$. 
At this fixed point $K_{s-} \to 1$, and  both perturbations are 
dimension $1$ operators. 
This system is invariant under the duality transformation $\phi_{s-} 
\leftrightarrow\theta_{s-}$. 
This model has been studied extensively in the literature~\cite{schulz,self}. 
It is equivalent to a theory of two Ising models. 
If the coupling constant in front of both operators is the same, as it
is the case in Eq.\  (\ref{eff1}), one of the Ising models is at its 
critical point. 
Equivalently, it can be regarded as a theory of two Majorana fermions, 
one of which is massive.
Hence this fixed point  is in the universality class of the two-dimensional
classical Ising model. 
The Ising order and disorder operators are given by $\sin \sqrt \pi \phi_{s-}$
and $\sin \sqrt \pi \theta_{s-}$ respectively. 
At this fixed point, both operators have scaling dimension $1/8$, as they
should at an Ising transition.
A small perturbation making $g_3\gtrsim g_4$ or $g_3 \lesssim  g_4$
causes a flow towards the CDW+SP or DSC fixed points respectively.
Thus $g_3=g_4>0$ is the phase boundary between  the phase CDW+SP and
a d-wave superconductor at $g_{s-}(0)=0$ and $g_{s+}(0)>0$.

\begin{figure}[h!]
\centering \epsfig{file=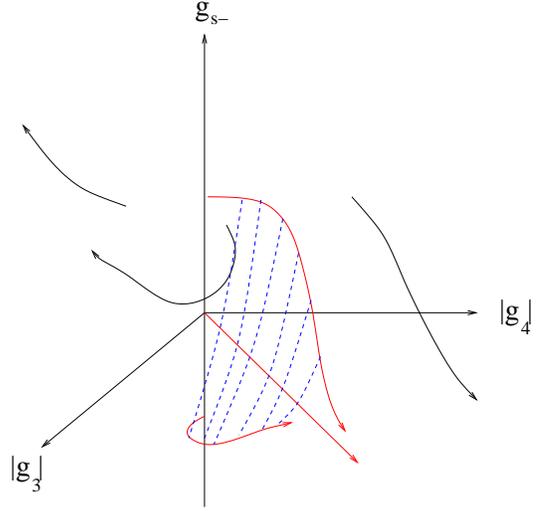,clip=,width=7cm,angle=0}
\caption{
RG flows in the 3D parameter space with $g_{s+}(0)>0$.
The dashed lines mark the critical surface.
$|g_3|$ wins over $|g_4|$ on the left of the surface,
and $|g_4|$ wins over $|g_3|$ on the right.
On the critical surface, the
RG trajectories flow to the line $|g_3|=|g_4|$.}
\label{flow}
\end{figure}

However, if the RG flows begin with $g_{s+}(0)<0$ along this direction, 
then the field $\phi_{s-}$ is no longer critical.
According to Eq. \ref {rg1}, in this regime $g_{s+}$ is marginally relevant 
and $g_{s+} \rightarrow-\infty$, with $g_3=g_4>0$ and $g_1=g_2<0$.
At this fixed point,  the fields  $\theta_{c-}$ and $\phi_{s+}$
acquire non-vanishing expectation values, and the residual interactions at 
this fixed point reduce to
\bea \label{eff2}
{\cal H}^2_{res}&=& \frac{\cos \sqrt{4\pi} \phi_{s-} }{ 2 (\pi a)^2 } 
( g_4^* \avg {\cos \sqrt{4\pi} \theta_{c-} } 
+g_1^* \avg{\cos \sqrt {4\pi} \phi_{s+} }) \nonumber \\
&+& \frac{\cos \sqrt{4\pi} \theta_{s-} }{2 (\pi a)^2 }   
(g_3^* \avg{\cos \sqrt {4\pi} \theta_{c-}}
 -g_2^* \avg {\cos \sqrt{4\pi} \phi_{s+}} ). 
\nonumber \\
&&
\eea
At this stage of RG, the  renormalized couplings
satisfy $g^*_4=g^*_3$ and $g_1^*=g_2^*$.
Once again we can take 
$\langle \cos(\sqrt{4\pi} \theta_{c-})\rangle=1$ (the renormalization of its
amplitude can be absorbed in a redefined coupling constant).
This effective theory has the same form as Eq.\ (\ref{eff1}). 
Hence this is also a theory of two Ising models.
However, unlike Eq.\  (\ref{eff1}) the amplitudes of the two dimension one 
operators are not equal. 
Hence, generically both Ising models are off-criticallity (or equivalently 
both species of Majorana fermions are massive). 
This corresponds to a finite correlation length and finite energy gap at
the phase boundary. 
Hence, in general this is a first order transition.
If $\avg {\phi_{s+}}=0$, then the term of $\theta_{s-}$ wins over that 
of $\phi_{s-}$  and $g_{s-}\rightarrow+\infty$ in the next step RG 
transformation. 
Conversely if $\avg {\phi_{s+}}=\sqrt \pi/2$, then the term of $\phi_{s-}$ 
wins over that of  $\theta_{s-}$ and $g_{s-}\rightarrow -\infty$
in the next step RG transformation.
Finally, RG  flows evolve to the CDW+SP  fixed point in the former case 
while towards the DSC fixed point in the latter.
Thus for $g_3=g_4>0$ and $g_{s+}(0)<0$, the phase transition at the boundary
of CDW+SP $\leftrightarrow$ DSC becomes first order as the correlation
length is now finite.
However, a second order transition is also possible here too. 
If the spin boson $\phi_{s+}$ is quantum disordered, then $\langle
\cos(\sqrt{4\pi}\phi_{s+})\rangle=0$ and once again we get an Ising
critical point  of the same kind discussed above. 
Hence the general conclusion is that this phase boundary may be at 
a second order transition (with Ising criticallity) or at a first order 
transition, with an Ising-like tricritical point in between.
Similarly,  $g_3=g_4<0$ at $g_{s-}=0$ is the boundary of DDW+DC 
$\leftrightarrow$ SSC, which is  critical and leads to the fixed point 
at $(0,0,-\infty,-\infty)$ or to a first order when $g_{s+}>0$ or
$g_{s+}<0$ respectively.

Another pair of fixed points $ (-\infty,0,\mp\infty,\pm\infty)$ control
the phase boundaries of the DDW+DC $\leftrightarrow$ DSC transition 
at $g_3=-g_4<0$, where $\langle \phi_{s+}\rangle=0$, and at the phase
boundary of the CDW+SP $\leftrightarrow$ SSC transition at $g_3=-g_4>0$,
where $\langle \phi_{s+}\rangle=\sqrt \pi/2$; $g_{s+}\rightarrow -\infty$ 
no matter what its initial value is.
The residual interaction for the  $s-$ sector is still described 
by Eq.\  (\ref{eff2}) but now with $g_3^*=-g_4^*$.
Thus, the amplitudes of $\cos \sqrt{4\pi} \phi_{s-}$ and
$\cos \sqrt{4\pi} \theta_{s-}$ are kept equal and this phase boundary 
is also in the universality class of the Ising critical point.
The Ising order and disorder operators can be determined accordingly.
The critical phase boundaries are summarized in Table \ref{boundary-quasi}.

The initial value $g_{s-}(0)$ has important effects on phase boundaries.  
In Fig.\ \ref{phase} we present the result of a numerical 
integration of Eq.\ (\ref{rg1}) for $g_{s+}(0)>0$;
$g_{s-}>0$ favors the growth of $|g_3|$ 
but disfavors that of $|g_4|$, and conversely $g_{s-}<0$ favors the 
growth of $|g_4|$ but disfavors that of $|g_3|$.
Let us begin with the case $g_{s-}(0)>0$.
For $|g_{3}(0)|\lesssim |g_{4}(0)|$, at first $g_{s-}$ decreases,
then it reaches a positive minimum and finally it increases.
Thus $|g_3|$ increases faster than $|g_4|$ and eventually it wins over it.
However, if $|g_{3}(0)|\ll |g_{4}(0)|$, $g_{s-}$ decreases monotonically 
to negative values and $|g_4|$ still wins over $|g_3|$.
As a result, both regions of the phase diagram with DDW+DC order and CDW+SP
order expand beyond the line $g_3=\pm g_4$, and the correspondingly areas
of d-wave and s-wave superconductivity correspondingly shrink.
Due to the symmetry of Eq.\  (\ref{rg1}), the situation is reversed 
for $g_{s-}(0)<0$.
For an initial point located on one of these phase boundaries, its RG
trajectory flows to the corresponding unstable fixed point,
as shown in Fig. \ref{flow}.
For $g_{s+}(0)<0$,  the effect of $g_{s-}(0)$ is similar,
but the phase boundaries CDW+SP $\leftrightarrow$ DSC and
DDW+DC $\leftrightarrow$ SSC are now first order transitions
and there are no accessible critical points.

We conclude this section with some comments on the DDW phase which has 
attracted considerable interest recently.
Until now there is  no solid numerical  evidence away from half-filling 
\cite{scala, chak2}. 
For the two-leg ladder, we find (see Appendix \ref{sec:ham}) that the 
DDW phase may exist but it is necessarily incommensurate. 
We also find $V_\perp$ large and positive reduces $g_4$ and enhances $g_{s-}$,
which is favorable for the DDW phase to exist.
However, a negative $g_3$ with magnitude comparable to $|g_4|$ is also needed.
Thus we suggest to look for it in the regimes $V_\perp \gg V_d \gg V_\pp>0$, 
which has only repulsive interactions, or for  $V_\perp>0>V_\pp$, 
which has some attractive interactions (and thus is less physically relevant).
These arguments agrees with the results of a recent two-dimensional mean-field 
calculation~\cite{tudor} that the Hubbard  $U$ alone can not stabilize the
DDW phase and that negative nearest-neighbor interactions are needed.
However,  $V_\perp, V_\pp<0$ together favor d-wave superconductivity over 
the DDW state.

\section{The Phase Diagram at half filling}
\label{sec:half}
Let us now discuss the phase diagram at half-filling. 
The main change is the presence of Umklapp terms. 
Compared to the incommensurate case discussed in Section \ref{sec:rg}, 
the main difference is that at half-filling the ${\mathbb Z}_2$ symmetries
behind two-fold degeneracies found in away from half-filling are now can
be spontaneously broken, with possible phase transitions between 
the CDW and the spin-Peierls phases, and between the DDW and the DC phases.
Since much of the analysis is rather similar, here we will only sketch the 
main differences.  

The set of RG equations are now more complicated:
\begin{eqnarray}
&&\frac{d K_{c+} }{ d l}= -\frac{K_{c+}^2 }{ 8 \pi^2} 
(g_{uc}^2+g_{u3}^2 +g_{u4}^2+g_{u5}^2)\nonumber \\
&&\frac{d K_{c-} }{ d l}=\frac{1 }{ 8 \pi^2} (g_3^2 +g_4^2+g_5^2)
\nonumber \\
&&\frac{d K_{s+} }{ d l}= -\frac{K_{s+}^2 }{ 8 \pi^2} 
(g_1^2 +g_2^2+g_5^2+g^2_{u5})
\nonumber\\
&&\frac{d K_{s-}}{ d l}=-\frac{K_{s-}^2 }{ 8 \pi^2} (g_1^2 +g_4^2+g_{u4}^2)+
\frac{1}{ 8 \pi^2} (g_2^2+g_3^2+g_{u3}^2) \nonumber \\
&&\frac{d g_1 }{ d l} = (2-K_{s+}-K_{s-}) g_1-\frac{g_4 g_5 }{ 2\pi}
-\frac{g_{u4} g_{u5} }{ 2\pi}
\nonumber \\
&&\frac{d g_2 }{ d l} = (2-K_{s+}-\frac{1}{K_{s-}}) g_2+\frac{g_3 g_5 }{ 2 \pi}
+\frac{g_{u3} g_{u5} }{ 2 \pi}
\nonumber \\
&&\frac{d g_3 }{ d l} = (2-\frac{1}{K_{c-}}-\frac{1}{K_{s-}}) g_3 
+\frac{g_2 g_5 }{ 2 \pi}+ \frac{g_{u3} g_{uc} }{ 2 \pi}
\nonumber \\ 
&&\frac{d g_4 }{ d l}= (2-\frac{1}{K_{c-}}-K_{s-}) g_4 -\frac{g_1 g_5 }{ 2 \pi}
+\frac{g_{u4} g_{uc} }{ 2 \pi} \nonumber \\
&&\frac{d g_5 }{ d l}= (2-\frac{1}{K_{c-}}-K_{s+}) g_5
-\frac{g_1 g_4 }{ 2\pi} +\frac{g_2 g_3 }{ 2 \pi}
+\frac{g_{u5} g_{uc} }{ 2 \pi}\nonumber \\
&&\frac{ d g_{uc} }{ d l}
= (2-K_{c+} -\frac{1}{K_{c-}}) g_{uc} +
\frac{g_3 g_{u3} }{ 2 \pi} +\frac{g_4 g_{u4} }{ 2 \pi}
+\frac{g_5 g_{u5} }{2 \pi} \nonumber \\
&&\frac{d g_{u3} }{ d l} = (2-K_{c+}-\frac{1}{K_{s-}}) g_{u3} 
+\frac{g_2 g_{u5} }{ 2 \pi}+
\frac{g_{3} g_{uc} }{ 2 \pi}\nonumber \\
&&\frac{d g_{u4} }{ d l} = (2-K_{c+}-K_{s-}) g_{u4} 
-\frac{g_1 g_{u5} }{ 2 \pi}+\frac{g_{4} g_{uc} }{ 2 \pi}\nonumber\\
&&\frac{d g_{u5} }{ d l}= (2-K_{c+}-K_{s+}) g_{u5}
-\frac{g_1 g_{u4} }{ 2\pi} +\frac{g_2 g_{u3} }{ 2 \pi}
+\frac{g_{5} g_{uc} }{ 2 \pi}. \nonumber\\
&&
\end{eqnarray}
We will not be interested here in solving these RG equations in their full
glory, but only in the regime where $K_{c+}\ll 1$ and $K_{c-}\sim 1$.
For this range of parameters there are a number of useful hierarchies of
scales which considerably simplify the analysis. 
 
Contrary to what happens away from half filling, the field $\phi_{c+}$ no 
longer decouples due to the effects of the Umklapp terms of 
Eq.\ \ref{HUmklapp}. 
Clearly, $\phi_{c+}$ plays a role quite similar to that of $\theta_{c-}$.
Indeed, in this regime, $g_{uc}$ is the most relevant coupling and it 
is associated to an operator with scaling dimension $K_{c+}+1/K_{c-}$.
This operator takes the RG flow close to a fixed point at which
the field $\phi_{c+}$ acquires a gap approximately of the form 
$m_{c+}\approx a^{-1} |g_{uc}(0)| ^{1/(1-K_{c+}(0)) }$.
In this regime the field $\theta_{c-}$ behaves roughly in the same way as 
in Eq.\ (\ref{rg1}). 
Here too the coupling constant $g_{c-}$ flows to strong coupling, 
$1/K_{c-}\rightarrow 0$ and a gap $m_{c-}$ develops in this sector as it
does away from half filling.  
We will set  $\avg{\phi_{c+}}=\sqrt \pi/2$, 
correspondingly $\avg{\theta_{c-}}=0$ or ${\sqrt \pi}$  
when $g_{uc}(0)>0$ or $<0$ respectively,
so that $\avg{\cos \sqrt{4\pi}\phi_{c+}}\approx -(a~m_{c+})^{K_{c+}(0)}$
and $\avg{ \cos \sqrt{4\pi}\theta_{c-} }\approx
\mbox{sgn}(g_{uc}) (a~m_{c-} )^{1/K_{c-}(0)}$.

Once the fields $\phi_{c+}$ and $\theta_{c-}$ become pinned close to their
classical values, the effective residual interactions among the remaining
fluctuating degrees of freedom have an effective Hamiltonian of the form
\begin{eqnarray}\label{eff}
{\cal H}_{eff}&=&\frac{\cos \sqrt{4\pi} \phi_{s+} }{ 2(\pi a)^2 }
(g_1 \cos \sqrt {4\pi} \phi_{s-}-g_2 \cos \sqrt{4\pi} \theta_{s-})
\nonumber \\
&+& \frac{1 }{ 2(\pi a)^2 }  
(g_3^*  \cos \sqrt {4\pi} \theta_{s-}
+g_4^*  \cos \sqrt{4\pi} \phi_{s-} \nonumber \\
&+& g_5^*  \cos \sqrt{4\pi} \phi_{s+}),
\end{eqnarray}
where
$g_{3,4}^*(0)= g_{3,4}(0) \langle \cos \sqrt{ 4 \pi} \theta_{c-} \rangle
-g_{u3,u4}(0)  \langle \cos \sqrt{ 4 \pi} \phi_{c+} \rangle$
and $g_5^*(0)=g_4^*(0)-g_3^*(0)$.
If $g_{uc}(0)$ is not small compared with the initial (or bare) values 
of the other coupling constants, this first step of the renormalization
group flow is rather quick. 
In this step the marginal coupling constants cannot not change very much
and thus $|\langle \cos \sqrt{ 4 \pi} \phi_{c+} \rangle| \gg |\langle 
\cos \sqrt{ 4 \pi} \theta_{c-} \rangle|$ is a good approximation.  
Hence the renormalized residual couplings are approximately  
$(g^*_3,g_4^*,g_5^*) \sim ( g_{u3}(0) ,g_{u4}(0) ,g_{u5}(0) ) $.

\begin{table}[h]
\begin{center}
\begin{tabular}{|c|c|c|c|c|c|c|c|}   \hline
 & $g_{uc}$  & $g_1,g_2$   & $g_3^*,g_4^*,g_5^*$        
& $\theta_{c-}$  & $\phi_{s+}$ & $(\phi_{s-},\theta_{s-})$&phase \\ \hline
1 &$+\infty$ & $0,-\infty$ &  $+\infty, 0, -\infty$ &
0 & 0              &(/,  $\frac{\sqrt\pi}{ 2}$) & SP \\ \hline
2 &$+\infty$   & $0, -\infty$ &  $-\infty, 0, +\infty$ 
& 0 & $\frac{\sqrt\pi}{ 2}$ & (/ , 0) & DDW \\ \hline
3 &$+\infty$ & $-\infty,0$  &  $0,+\infty,+\infty$ &
0 & $\frac{\sqrt\pi}{ 2}$ & ($\frac{\sqrt\pi }{ 2}$,  /) &DSC+SDW \\ \hline 
4 &$+\infty$ & $-\infty,0$  &  $0,-\infty,-\infty$ &
0 & 0             & (0, /) &SSC+DC$^t$ \\ \hline
5 &$-\infty$ & $0, -\infty$ &  $-\infty, 0, +\infty$ &  
$\frac{\sqrt\pi}{ 2}$ & $ \frac{\sqrt\pi}{ 2}$ &(/ ,0) & CDW  \\ \hline
6 & $-\infty$ & $0, -\infty$& $+\infty, 0, -\infty$ &
$\frac{\sqrt\pi}{ 2}$ & 0      &(/, $\frac{\sqrt\pi}{ 2}$) &DC \\ \hline
7 &$-\infty$ & $-\infty,0$  &  $0,-\infty,-\infty$ &
$\frac{\sqrt \pi }{ 2}$ & 0             & (0 , /)&DSC+SP$^t$ \\ \hline
8 &$-\infty$ & $-\infty,0$  &  $0,+\infty,+\infty$ &
$\frac{\sqrt\pi}{ 2}$ & $\frac{\sqrt \pi}{ 2}$ 
& $(\frac{\sqrt \pi }{ 2},  /)$ &SSC+DDW$^t$\\ \hline  
\end{tabular} 
\end{center}
\caption{Fixed points at half-filling: Stable fixed points and 
corresponding gapped phases.
We have set $\langle \phi_{c+}\rangle=\sqrt \pi/2$.
The $SU(2)$ condition requires $g_5^*=g_4^*-g_3^*$. 
Phases 1,2,5,6 have true Ising type long range order,
while 3,4,7,8 are  quantum disordered Haldane-like phases .}
\label{table:Umklapp-stable}
\end{table}

\begin{table}
\vspace{5mm}
\begin{center}
\begin{tabular}{|c|c|c|c|c|c|c|} \hline
 &$g_{uc}$ & $g_{s+}$ & $g_3^*, g_4^*, g_5^*$ &$\theta_{c-} $ &$\phi_{s+}$
& Transition  \\ \hline
1 &$+\infty$& 0 & $+\infty,+\infty,0$  
&0 & / & DSC+SDW$\leftrightarrow$ SP \\ \hline  
2 &$+\infty$& 0& $-\infty,-\infty,0$  
&0 & / &SSC+DC$^t$ $\leftrightarrow$ DDW  \\ \hline 
3 &$+\infty$&$ -\infty$& $-\infty,+\infty,+\infty$  
&0 &$\frac{\sqrt \pi}{ 2}$  
&DSC+SDW $\leftrightarrow$ DDW  \\ \hline 
4 &$+\infty$&$ -\infty$& $+\infty,-\infty,-\infty$  
&0 & 0  &SSC+DC$^t$ $\leftrightarrow$ SP  \\ \hline 
5 & $-\infty$& 0 &  $-\infty,-\infty,0$ 
& $\frac{\sqrt \pi }{ 2} $ & / &
DSC+SP$^t$ $\leftrightarrow$ CDW  \\ \hline
6 & $-\infty$& 0 &  $+\infty,+\infty,0$ 
& $\frac{\sqrt \pi }{ 2} $ & / &
SSC+DDW$^t$ $\leftrightarrow$ DC \\ \hline
7 &$-\infty$&$ -\infty $& $+\infty,-\infty,-\infty$  
&$\frac{\sqrt \pi }{ 2}$ & 0  &
DSC+SP$^t$  $\leftrightarrow$ DC
\\ \hline
8 &$-\infty$&$ -\infty $& $-\infty,+\infty,+\infty$  
&$\frac{\sqrt \pi }{ 2}$ &$ \frac{\sqrt \pi }{ 2}$  &
SSC+DDW$^t$   $\leftrightarrow$ CDW \\ \hline 
\end{tabular} 
\end{center}
\caption{Fixed points at half-filling: Unstable fixed points which have 
the common fixed value $g_{s-}=0$.
Here too $\langle \phi_{c+}\rangle=\sqrt \pi/2$, and the $SU(2)$ condition 
requires $g_5^*=g_4^*-g_3^*$.
The column on the right indicates which transition is controlled by each
unstable fixed point.}
\label{table:Umklapp-unstable}
\end{table}

\begin{figure}
\centering \epsfig{file=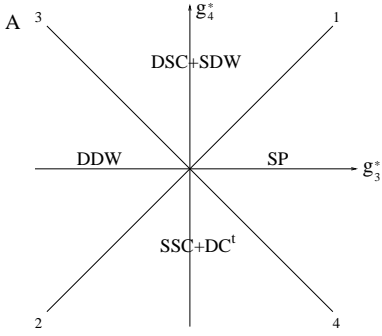,clip=,width=5cm,angle=0}
\caption{Stable phases and phase boundaries at half-filling 
with $g_{s_-}(0)=0,g_{s+}(0)>0$, and  $g_{uc}(0)>0$.
Phase boundaries 1,2,5,6 are represent first order transitions when
$g_{s+}(0)<0$. The critical fixed points for the transitions from phases
in this figure to their counterparts in Fig.\ \ref{phaseUmklapp2} are 
analogous to those of Fig. \ref{phase}.
}
\label{phaseUmklapp1}
\end{figure}

\begin{figure}
\centering \epsfig{file=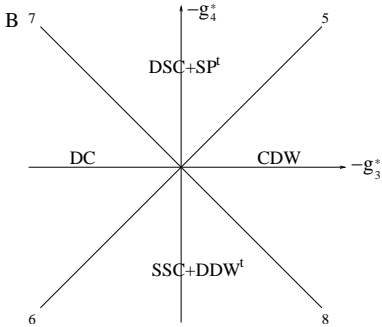,clip=,width=5cm,angle=0}
\caption{
Phase diagram and boundaries at half-filling 
with $g_{s_-}(0)=0,g_{s+}(0)>0$ and 
$g_{uc}(0)<0$.
Notice that here we use $-g_3^*$ and $-g_4^*$ as the $x,y$ axes. 
}
\label{phaseUmklapp2}
\end{figure}

The new RG equations, which control the subsequent RG flow,  are the same 
as in Eq.\ (\ref{rg1}) after setting  $1/K_{c-} \to 0$.
Here too the $SU(2)$ condition $g_5^*=g_4^*-g_3^*$ is obeyed, albeit 
among renormalized couplings.
The resulting stable phases and the phase boundaries between them are
given in the phase diagrams of Fig. \ref{phaseUmklapp1} and 
\ref{phaseUmklapp2}. 
The corresponding stable fixed points and values of pinned fields are 
summarized in Table \ref{table:Umklapp-stable}. 
The critical (or unstable) fixed points are given in Table 
\ref{table:Umklapp-unstable}.
Umklapp terms break the symmetry group to $C_{4v}$ and thus remove the 
degeneracy between CDW and SP phases, and between the DDW and DC phases.
Hence, all four states become distinct phases with true long range order,
which break the  residual  ${\mathbb Z}_2$ symmetry  spontaneously.
At the quantum phase transitions between CDW and SP, and between
DDW and DC, the symmetry is U(1).

Perturbative RG studies of Refs.\  \cite{lin} and \cite{fjarestad} have
described the CDW  and DDW fixed points with the property that the 
coupling constants (written in our notation) satisfy
\begin{eqnarray}\label{purepert}
-g_2&=&\pm g_3=\pm g_5=-g_{u3}=g_{u5}=
\mp g_{uc}\rightarrow +\infty \nonumber \\
g_1&=&g_4=g_{u4}=0.
\end{eqnarray}
where the upper (lower) sign holds for the CDW (DDW) phase. 
It turns out that a model with this particular choice of coupling constants 
was been proposed by Scalapino, Zhang and Hanke (SZH) ~\cite{so5-ladder} as
a ladder model of the $SO(5)$ theory. 
However, Lin, Balents and Fisher found that, at least to one loop order
in a perturbative RG~\cite{lin}, the symmetry is enlarged actually to $SO(8)$.
Moreover these authors found, also within a perturbative RG, that the $SO(8)$
manifold is at least locally stable, \textit{i.\ e.\/}  small deviations
from this trajectory converge to this trajectory under the RG flow. 
Interestingly, the $SO(8)$ manifold is an integrable fermionic system for 
which a number of exact properties have been calculated using the Bethe
Ansatz~\cite{konik}. 
$SO(8)$ is clearly a dynamical symmetry which is possible because the 
operators that are involved (back in the fermionic representation) are 
all dimension 2, they are superficially marginal 
but become marginally relevant due to fluctuations leading to the 
development of a gap.

However, for more generic values of the coupling constants this dynamical 
symmetry does not necessarily arise.
It is not know how large the basin of attraction of the $SO(8)$ manifold
actually is. 
In fact using bosonization methods we find that far away from the $SO(8)$ 
manifold the scaling dimensions of these operators  begin to differ 
significantly from each other and thus evolve differently under the RG
(see Eq.\  (\ref{purepert})). 
In particular, by checking their scaling dimensions, we find that the 
renormalized couplings can renormalize so differently from each other 
as 
\bea
&&|g_2|\ll |g_3|=|g_5|\ll |g_{u3}| =|g_{u_4}| \ll |g_c|\rightarrow \infty
\nonumber\\
&& g_1, g_4, g_{u4}\rightarrow 0
\eea
in all the four phases of CDW, DDW, SP and DC ( Recall that the 
signs of the coupling constants change in some of the phases. ).
Nevertheless, what is clear is that the spectrum found in these more
anisotropic (and more generic) regimes is smoothly connected to the multiplets
found in the $SO(8)$ limit. In other words there is no phase transition
separating these regimes, but the spectrum is organized differently.

Let us now discuss the phase transitions between the CDW and SP phases, and
between the DDW and DC and phases, and to the associated critical fixed points.
As we noted before, these  phase transitions are driven by the
Umklapp terms, the most relevant of which is controlled by the coupling 
constant $g_{uc}$. 
Hence, {\sl at the critical point} separating the SP and CDW phases, 
and between the DDW abd DC phases, the Umklapp terms are tuned to zero. 
The critical fixed points coincide with the stable fixed points 
of the incommensurate CDW+SP phase and DDW+DC phase respectively. 
In both cases the transition is controlled by the sign of $g_{uc}$.
We also note that the renormalized coupling constant $g_3^*$ has different 
signs on both sides of this phase transition.
This is because, close to the transition  $g_3^*\approx g_3\avg{\cos 
\sqrt {4\pi} \theta_{c-}}$,  and $\avg{\theta_{c-}}=0$ in the SP phase
while $\avg{\theta_{c-}}= \sqrt \pi/2$ in the CDW phase.
The same is true for the phase transition between the DDW and DC phases.

It can be shown that, if only charge interactions are considered~\cite{gcg3},
then $g_{uc}=g_{u3}$ at the bare level.
In this regime the CDW and SP phases are more easily accessible than 
the DDW and DC phases.
There is strong numerical evidence for a commensurate DDW phase at
half-filling in a t-J-Hubbard ladder \cite{marston} who included 
Heisenberg-like exchange interactions at the microscopic level.
It is easy to see that although the inclusion of microscopic exchange 
interactions does not lead to a different low energy theory,
it changes the strengths of the different effective couplings.
In particular it makes the DDW phase more accessible.
For simplicity, we discuss the conditions of the commensurate DDW phase
on the SZH ladder which only includes nonzero interactions
$U, V_\perp, J_\perp$.
The coupling constants are given in the weak interaction limit
in Appendix A. Let us suppose that $V_\perp$ and $J_\perp>0$.
First of all, we need positive $g_{c+}$ to set up the overall repulsive
interaction i.e., $U+2 V_\perp>0$.
A large $J_\perp$ helps to make $g_{uc}>0$ and $g_{u3}<0$ 
simultaneously {\it i. e.} , $\frac{1}{4}J_\perp> U-V_\perp> -\frac{3}{4} J_\perp$.
But $J_\perp$ can not be too large, otherwise negative $g_{s-}$ 
suppresses the DDW phase. 
For $|g_{u3}|>|g_{u4}|$, which can be achieved with $U<0$, this phase is
stabilized. 
But $|U|$ can not be too large, otherwise $g_{c+}$ would become negative.
The region where the commensurate DDW was found in Ref.\ \cite{marston}
agrees with this analysis.
Again, we need to keep in mind that this naive analysis only makes sense
in the weak coupling limit, which also neglects effects from many
irrelevant operators. Thus, we do not expect this analysis to give a precise
location of the phase boundary.

Now we discuss the remaining phases and phase transitions.
Upon a careful study of which fields become pinned and what are their 
allowed expectation values,  we conclude that the remaining four phases are 
actually quantum disordered Haldane-like  states.
For example, there is a phase in which $d$-wave superconductivity
and the SDW order parameters (DSC+SDW) are quantum disordered. 
The order parameter for DSC is very sensitive to fluctuations in the $c+$
sector since $O_{DSC}\propto e^{i\theta_{c+}}$. 
Similarly, the $x$, $y$ and $z$ components of the SDW order parameter 
are controlled by fluctuations in the $s\pm$ sector since  
$\vec O_{SDW}\propto \left(\sin(\sqrt{\pi}\theta_{s-}),\sin(\sqrt{\pi}
\theta_{s+}),\cos(\sqrt{\pi}\theta_{s+})\right)$.
At this fixed point the fields $\theta_{c+}$ and $\theta_{s\pm}$ 
are not pinned and fluctuate wildly.
Nevertheless, the remaining fields in the expressions for these 
order parameters do provide for a finite amplitude even though the  
fluctuations of both phase and orientation are so strong that the
system is quantum disordered.
The analysis of other three phases, $s$-wave superconductivity and triplet DC
(SSC+DC$^t$), $d$-wave superconductor and triplet spin-Peierls (DSC+SP$^t$), 
and $s$-wave superconductor and triplet $d$-density wave (SSC+DDW$^t$), 
is similar.
Because of  large charge gaps, the low energy physics of their spin sector,
may be described by the corresponding $O(3)$ non-linear $\sigma$ model 
without a Berry phase term, which is quantum disordered.

The phase transition between the DSC+SDW phase and the DSC+SP$^t$ phase 
(see Fig.\ \ref{phaseUmklapp1} and Fig.\ \ref{phaseUmklapp2}) is the
commensurate limit of the $d$-wave superconductor found away from half filling.
A similar relation holds for the phase transition between the SSC+DC$^t$ phase,
the SSC+DDW$^t$ phase and the $s$-wave superconductor.

Finally, let us discuss the unstable fixed points with $|g_3^*|=|g_4^*|
\rightarrow \infty, g_{s-}=0$, summarized in Table 
\ref{table:Umklapp-unstable}. 
The RG flows starting from the phase boundaries  with $g_{s+}(0)>0$ evolve
towards these fixed points. 
At these phase  boundaries, the order parameters for CDW, SP, DC and
DDW have power-law correlations and have  scaling dimension 3/8 at the
fixed points denoted by 1,2,5, and 6, and scaling dimension 1/8 at the 
fixed points denoted by 3, 4, 7 and 8 (see  Figs.\ \ref{phaseUmklapp1} 
and \ref{phaseUmklapp2}). On these phase boundaries the $d$-wave and
$s$-wave superconducting order parameters are quantum disordered.
Similarly, the SDW, SP$^t$, DC$^t$ and DDW$^t$ order parameters have 
power-law correlations and their scaling dimension is 3/8 at the points
1,2,5,6 but are quantum disordered at points 3,4,7,8. 
For $g_{s-}(0)=0$, at these phase boundaries the renormalized couplings 
satisfy $|g_3^*|=|g_4^*|$ as before. 
Nonzero $g_{s-}(0)$ also has similar effects on these phase boundaries:
$g_{s-}(0)>(<)0$ favors phases CDW, SP, DC and DDW (DSC+SDW, SSC+DC$^t$, 
DSC+SP$^t$ and SSC+DDW$^t$ ) respectively.
When $g_{s+}(0)<0$,  the situation is  similar except that transitions 
1,2,5,6 become the first order and there are no corresponding unstable 
fixed points.


\section{Conclusions}
\label{sec:conclusions}

In summary, in this paper we studied the problem of competing orders in 
two-leg ladders, which were mapped to two-coupled Luttinger liquids
with p-h symmetry at both low doping and at  half-filling.
We used (Abelian) bosonization and RG  methods to study the phase 
diagrams of these ladders both at half filling and at low doping.
Stable and unstable fixed points  of the RG flows with the corresponding
phases  and phase boundaries were investigated in detail.
First order transitions when $g_{s+}(0)<0$ are found and the effects of
$g_{s-}(0)$ to phase boundaries are discussed.
The $C_{\infty v}$ symmetry makes  CDW and spin-Peierls, DC and DDW 
degenerate. 
In the absence of Umklapp terms there is incommensurate quasi long 
range order. These degeneracies are removed at half-filling where true
long range order appears.
Power law fluctuating $d$-wave and $s$-wave supreconducting phases
at low doping levels become quantum disordered at half-filling, 
with finite amplitudes among DSC, SSC and SDW, DC$^t$, SP$^t$,  DDW$^t$ 
respectively. 
Suggestions on how to best find these phases in numerical simulations
were given.

After this paper was submitted for publication, we became aware of the work by 
Tsuchiizu and Furusaki on a very similar model (at half-filling) \cite{tsuchiizu}.
In this work these authors also obtained the same eight insulating phases we
found here at half-filling.
Also after this work was submitted, we learned of the numerical work by Schollwock 
{\it et. al.} \cite{schollwock}
on a DMRG study of a similar ladder model away from the half
filling. 
At low doping these authors found that their results are consistent with
an inhomogeneous picture of the doped state in which the system is locally
commensurate.
It is our understanding that at long length scales the system is
actually incommensurate with discommensurations (or kinks) separating the
locally commensurate regions. On length scales long compared to the distance
between kinks, this state behaves like an effective ``elastic solid" which in
one-dimension has the same quantum critical behavior as a Luttinger liquid. 
Thus, this state is qualitatively equivalent to our
weak coupling picture, albeit with substantially renormalized parameters.

\begin{acknowledgments}

We thank S. Chakravarty  and P. Phillips for helpful discussions. 
This work is supported  by NSF grant
DMR98-17941 and DMR01-32990 at UIUC.   
WVL is also supported  in part by funds
provided by the U.S. Department of Energy (DOE) under cooperative
research agreement \#DF-FC02-94ER40818 at MIT.

\end{acknowledgments}

\begin{widetext}
\appendix
\section{ Fermionic Hamiltonian}
\label{sec:ham}
We considered an extended Hubbard model on a ladder with a Hamiltonian 
of the following form
\begin{eqnarray}
 H&=&-t \sum_{\langle i,j \sigma \rangle} 
\big\{c^\dagger_{i,j\sigma} c_{i+1,j\sigma}+h.c.\big\}
-t_\perp \sum_{\langle i \sigma \rangle} 
\big \{c^\dagger_{i,0\sigma} c_{i,1\sigma}+h.c.\big\}
+ U\sum_{i,j} n_{i,j\uparrow}n_{i,j\downarrow} 
+ V_\pp\sum_{i,j} n_{i,j}n_{i+1,j}\nonumber \\
&+&V_\perp\sum_{i} n_{1,i}n_{2,i} 
+V_{d}\sum_i(n_{i,1}n_{i+1,2}+n_{i,2}n_{i+1,1} )+J_\perp
\sum_i \vec{S}_{i,1}\cdot \vec{S}_{i,2}
+J_\pp\sum_{ij} \vec{S}_{i,j} \cdot \vec{S}_{i+1,j}
\end{eqnarray}
Here $i$ labels the sites along legs and $j$ labels the legs (or rungs); the
coupling constants $U$, $V_\pp,V_\perp$ and $V_d$ represent the on-site 
Hubbard interaction and various nearest and next-nearest neighbor Coulomb 
interactions, and $J_\perp$ and $J_\pp$ are the Heisenberg interaction
along the rungs and chains respectively. 

After diagonalizing the kinetic part, we can rewrite the above Hamiltonian with
the right and left movers of the bonding and anti-bonding bands represented by 
the operators $\psi_{R1},\psi_{L1}, \psi_{R2},\psi_{L2}$ as below, where
$\psi_{1,2}(x)=( c_{i1}\pm c_{i2})/\sqrt {2 a}$. 
In the low energy limit, the free part of the continuum Hamiltonian density 
can be written as
\begin{eqnarray}
{\cal H}_0&=&  v_{f1}\big\{ {\pi\over2} (J_{L1}J_{L1} +J_{R1} J_{R1}  )+
\frac{2}{3}\pi (\vec{J}_{L1}\cdot \vec{J}_{L1}  +\vec{J}_{R1}\cdot 
\vec{J}_{R1} ) \big\}
\nonumber\\
&+& v_{f2}\big\{ {\pi\over2} (J_{L2}J_{L2} +J_{R2} J_{R2}  )+
\frac{2}{3}\pi (\vec{J}_{L2}\cdot \vec{J}_{L2}  +\vec{J}_{R2}\cdot 
\vec{J}_{R2} ) \big\},
\end{eqnarray}
where  $J_{i,R,L}=\psi^\dagger_{i,R,L\sigma}\psi_{i,R,L\sigma}$, and 
$\vec{J}_{i,R,L}=\psi^\dagger_{i,R,L\sigma}\psi_{i,R,L\sigma}$, are the 
right and left moving components of the charge and spin current density
for the bonding ($i=1$) and anti-bonding ($i=2$) fermions.

The interaction part of the Hamiltonian splits into several terms. 
First, we have a set of  terms involving only the 
charge currents:
\begin{eqnarray}
{\cal H}_{int,c}&=& \Big \{ \frac{U}{ 8} +\frac{1}{ 2} (V_\pp+V_d) 
+ \frac{V_\perp }{ 4}
\Big \}(J_{1R}J_{1R}+J_{1L}J_{1L}+J_{2R}J_{2R}+J_{2L}J_{2L}) \nonumber \\
&+&
\Big \{ \frac{U}{ 4} +(V_\pp+V_d)(1- \frac{\cos 2k_{f1} }{ 2} )+
\frac{V_\perp }{ 4} -\frac{3}{8} \cos 2 k_{f1} J_\pp -\frac{3}{16} J_\perp
\Big  \} J_{1R}J_{1L} \nonumber \\
&+&\Big \{ \frac{U}{ 4} +(V_\pp+V_d)(1- \frac{\cos 2k_{f2} }{ 2} )
+\frac{V_\perp }{ 4}  -\frac{3}{8} \cos 2 k_{f2} J_\pp -\frac{3}{16} J_\perp
\Big \}J_{2R}J_{2L}\nonumber\\
&+& \Big \{  \frac{U}{ 4} +V_\pp(1-\frac{1}{2}\cos k_- )+
V_d(1+ \frac{1}{2} \cos k_- )+\frac{3}{ 4} V_\perp 
-\frac{3}{8} J_\pp \cos k_-  +\frac{3} {16} J_\perp
\Big \} (J_{1R}J_{2R}+J_{1L}J_{2L}) \nonumber \\
&+& \Big \{ \frac{U}{4} +V_\pp ( 1-\frac{1}{2}\cos k+ ) 
+ V_d (1+\frac{1}{2} \cos k_+) +\frac{3}{ 4} V_\perp 
-\frac{3}{8} J_\pp \cos k_+  +\frac{3} {16} J_\perp
\Big \} (J_{1R}J_{2L}+J_{1L}J_{2R}),
\end{eqnarray}
where $k_+ =k_{f1}+k_{f2}=\pi(1-\delta)$,~ $k_-=k_{f1}-k_{f2}= 2 \sin^{-1} 
\big[ t_\perp/ (2 \cos {\pi\delta \over 2}) \big]$.

Next we have the couplings involving the spin currents:
\begin{eqnarray}
{\cal H}_{int,s}&=& \big\{-\frac{U}{ 6} +\frac{J_\pp}{2} +\frac{J_\perp}{4}
 \big\}
({\vec J}_{1R}{\vec J}_{1R}+{\vec J}_{1L}{\vec J}_{1L}+
{\vec J}_{2R}{\vec J}_{2R}+{\vec J}_{2L}{\vec J}_{2L}) 
\nonumber \\
&-&\Big \{U+2(V_\pp+V_d) \cos2k_{f1}+V_\perp -\frac{3}{4} J_\perp \Big \}
{\vec J}_{1R}{\vec J}_{1L}
-\Big \{ U+2(V_\pp+V_d) \cos2k_{f2}+V_\perp -\frac{3}{4} J_\perp  \Big \}
{\vec J}_{2R}{\vec J}_{2L}\nonumber\\
&-&\Big \{ U+2(V_\pp-V_d)\cos k_- -V_\perp 
-(1+\frac{1}{2}\cos k_-) J_\pp-\frac{J_\perp}{4}
\Big \}
( {\vec J}_{1R}{\vec J}_{2R}+{\vec J}_{1L}{\vec J}_{2L} )\nonumber\\
&-& \Big \{U+2(V_\pp-V_d)\cos k_+ -V_\perp 
-(1+\frac{1}{2}\cos k_+) J_\pp-\frac{J_\perp}{4}
\Big \} 
( {\vec J}_{1R}{\vec J}_{2L}+{\vec J}_{1L}{\vec J}_{2R} ).
\end{eqnarray}

Next we have the low energy couplings associated with singlet-pair and
triplet-pair tunneling:
\begin{eqnarray}
{\cal H}_{int,pt}&=&\Big \{U+ (2(V_\pp-V_d)-\frac{3}{2} J_\pp)
\cos k_{f1} \cos k_{f2} -V_\perp +\frac{3}{4} J_\perp
\Big \}(\Delta^\dagger_1 \Delta_2+h.c.) \nonumber \\
&+&\Big \{2(V_\pp-V_d )+\frac{J_\pp}{2}\Big\} \sin k_{f1} \sin k_{f2} ~
(\vec{\Delta}^\dagger_1 \vec{\Delta}_2+h.c.)
\end{eqnarray}
 where
$\Delta= (\psi_{R\uparrow}\psi_{L\downarrow}-\psi_{R\downarrow}\psi_{L\uparrow}
 )/\sqrt{2}$ is the singlet pair operator on a given chain and $\vec{\Delta}$ 
is its triplet counterpart. 
Notice that $1$ and $2$ stand here for the chain label.

Finally, the low energy Umklapp scattering terms are
\begin{eqnarray}
{\cal H}_{um}&=& e^{ 2i\delta\pi x } \Big \{ \big \{
\frac{U}{ 4} +e^{i\delta\pi} \big [  V_\pp(\frac{1}{ 2} -\cos k_- )
-V_d (\frac{1}{ 2} +\cos k_-) +\frac{3}{8} J_\pp  \big ] +\frac{3}{ 4} V_\perp 
+\frac{3}{16} J_\perp \big \}
N^\dagger_1N^\dagger_2 \nonumber\\
&-& \big \{U+e^{i\delta\pi} ( -2(V_\pp-V_d)+(\frac{1}{2} +\cos k_-) J_\pp)
-V_\perp -\frac{J_\perp}{4}
\big \}
\vec{N}^\dagger_1\vec{N}^\dagger_2  \nonumber\\
&+& \big \{ \frac{U}{ 2} -(V_\pp-V_d-\frac{3}{4} J_\pp
) e^{i\delta \pi } -\frac{V_\perp }{ 2}+\frac{3}{8} J_\perp
\big \}
(m^\dagger_{1R}m_{2L} +m^\dagger_{2R}m_{1L} ) \Big \}
+h.c,
\end{eqnarray}
Here 
$N^\dagger=\psi^\dagger_{R\sigma}\psi_{L\sigma}$
and  $\vec{N}^\dagger= \psi^\dagger_{R\sigma}(\vec{\sigma}/2)\psi_{L\sigma}$
are  CDW and SDW (N\'eel) order parameters, respectively.
$m$ is the paring order with $2k_f$ momentum, for example,
$m_R=\psi_{R,\uparrow}\psi_{R,\downarrow}$.

Following the standard Bosonization procedure with the assumption of Eq. 
\ref{ph}, we arrive at the bosonized Hamiltonian density in the section II. 
The bare values of the weak coupling constants are given as
\bea
g_{c+}&=&  U+ V_\pp\big[4+\cos \pi\delta ~(1+\cos k_{f-}) \big] +2~V_\perp
+ V_d \big[ 4-\cos \pi \delta ~ (1-\cos k_{f-}) \big]
+\frac{3}{4} J_\pp \cos \pi\delta(1+\cos k_-) ,
\nonumber \\
g_{c-}&=& -(V_\pp +\frac{3}{4} J_\pp) \cos \pi\delta~(1-\cos k_{f-}) 
- V_\perp +\cos \pi\delta~(1+\cos k_{f-} ) V_d-\frac{3}{4}J_\perp,
\nonumber\\
g_{s+}&=& U -V_\pp~ \cos \pi \delta~ (1+\cos k_{f-} )
+V_d \cos \pi \delta~ (1-\cos k_{f-}) 
-\frac{J_\pp}{2} (1-\frac{1}{2}\cos\pi\delta)
-\frac{J_\perp}{2}
, \nonumber \\
g_{s-}&=& V_\pp~ \cos \pi \delta~ (1-\cos k_{f-} )+V_\perp
-V_d \cos \pi \delta~ (1+\cos k_{f-})
+\frac{J_\pp}{2} (1-\frac{1}{2}\cos\pi\delta)-\frac{1}{4} J_\perp,
\nonumber \\
g_3&=&2 (V_\pp-V_d+\frac{J_\pp}{4})
\big[ \cos k_- + \cos \pi\delta \big],\nonumber\\
g_4&=& U+ 2 (V_\pp-V_d)  \cos k_- -V_\perp+ J_\pp(\cos\pi\delta-\frac{1}{2}
\cos k_-) +\frac{3}{4} J_\perp, \nonumber \\
g_{uc}&=& U-2 \cos\pi\delta~(V_\pp-V_d-\frac{3}{2} J_\pp)
-V_\perp+\frac{3}{4} J_\perp,\nonumber\\
g_{u3}&=& U-2\cos\pi\delta~ 
\big [V_\pp-V_d-J_\pp (\frac{1}{4}+\frac{1}{2} \cos k_-)\big ]-V_\perp
-\frac{J_\perp}{4},
\nonumber \\
g_{u_4}&=& U-2\cos \pi\delta~\big[(V_\pp+V_d)\cos k_- -J_\pp (1+\frac{1}{2} 
\cos k_-) \big] +V_\perp+\frac{J_\perp}{4},\nonumber\\
g_1&=& g_{s+} +g_{s-}, \hspace{10mm} g_2=  g_{s+} -g_{s-}, \hspace{10mm}
g_5=g_4-g_3, \hspace{10mm} g_{u5}=g_{u_4}-g_{u3},
\eea
where $\cos k_-= 1- t_\perp^2/(2t^2)$. 
Up to the first order, these coupling constants are independent on the 
doping $\delta$.

When away from the half-filling, the particle-hole symmetry Eq. \ref{ph}
only holds approximately at small doping $\delta$ as $k_+-\pi=\delta~\pi, 
\Delta v_f/a= \delta ~t_\perp \pi$.
Taking these into account, there are some small residue terms as appearing
in Eq. \ref{broken-ph1}, \ref{broken-ph2}, \ref{broken-ph3}, 
they vanish linearly with doping.
The corresponding coupling constants are
\bea
\Delta g_c&=& \{ \frac{1}{2} (V_\pp+V_d)+\frac{3}{8} J_\pp \}
\sin  \pi\delta ~\sin k_{f-}, \hspace{3cm}
\Delta g_s= -\frac{1}{2} (V_\pp+V_d)  
\sin \pi \delta~ \sin k_{f-}, \nonumber\\
\Delta g_{uc}&=& -2 \sin \pi \delta ~(V_\pp-V_d-\frac{3}{4}J_\pp)
,  \hspace{4.2cm}
\Delta g_{u3}=-2 \sin \pi \delta ~\big[V_\pp-V_d
- J_\pp (\frac{1}{4}+\frac{\cos k_-}{2}) \big]  \nonumber\\
\Delta g_{u4}&=&-2 \sin \pi \delta~ 
\big[ (V_\pp+V_d) \cos k_- -J_\pp (1+\frac{1}{2} \cos k_-) \big]
, \hspace{1cm}
\Delta g_{u5}=\Delta g_{u3}-\Delta g_{u4}.
\eea

\section{Bosonic representation of the order parameters}
\label{sec:order-p}

The difference of the charge density  between two legs reads
$\sum_\sigma (-)^{j+1}  c^\dagger_{j\sigma}(i) c_{j\sigma}(i)=\sum_\sigma 
\psi_{1\sigma}^\dagger(x)\psi_{2\sigma}(x)
+ \psi_{2\sigma}^\dagger(x)\psi_{1\sigma}(x)$.
After expressed by the right and left movers,
it contains the staggered part, ie. $O_{CDW}$.
Similar situation happens to its triplet counterpart $O_{SDW,z,x,y}$.
Using the bosonization identities:
$\psi_{R,L}(x)=1/\sqrt{ 2\pi a}~ 
\exp\{\pm i{\sqrt \pi} (\phi(x)\pm \theta(x))\}$ and
we can obtain their bosonic expressions as below,
\begin{eqnarray}
\left. \begin{array}{r}
O_{CDW}(x)\\
\vec{O}_{SDW,z,x,y}(x)\\
\end{array}\right\}
&=&
(-)^x  e^{-i\delta\pi x} 
\left \{ \begin{array}{c}
\psi^\dagger_{1L\sigma}(x)\psi_{2R\sigma}(x)+
\psi^\dagger_{2L\sigma}(x)\psi_{1R\sigma}(x)  +h.c.\\
\psi^\dagger_{1L\alpha}(x)( {\vec\sigma/ 2})_{\alpha\beta}
\psi_{2R\beta}(x)+
\psi^\dagger_{2L\alpha}(x) ( {\vec\sigma/ 2})_{\alpha\beta}
\psi_{1R\beta}(x) +h.c.\nonumber\\
\end{array}\right. \nonumber\\
&\propto& \frac{2 \Gamma}{ \pi a}
\big \{ \cos (\sqrt{\pi} \phi_{c+} -\delta \pi x)
\left\{\begin{array}{r}
2 \cos \sqrt{\pi} \theta_{c-}
\cos \sqrt{\pi} \phi_{s+} \sin \sqrt{\pi}\theta_{s-} \\
\sin \sqrt{\pi} \theta_{c-}
\left\{\begin{array}{r}
\cos \sqrt{\pi}  \phi_{s+}   \cos \sqrt{\pi}\theta_{s-}\\
\cos \sqrt{\pi}  \theta_{s+} \cos \sqrt{\pi}\phi_{s-}  \\
-\sin \sqrt{\pi}  \theta_{s+} \cos \sqrt{\pi}\phi_{s-} \\
\end{array}\right. \\
\end{array}\right. \nonumber\\
&&  +
 \sin( \sqrt{\pi}  \phi_{c+}-\delta \pi x) 
\left \{ \begin{array}{r}
-2 \sin \sqrt{\pi}\theta_{c-}
\sin \sqrt{\pi} \phi_{s+} \cos \sqrt{\pi} \theta_{s-} \\
-\cos \sqrt{\pi} \theta{c-}
\left \{ \begin{array}{r}
\sin \sqrt{\pi} \phi_{s+}   \sin \sqrt{\pi} \theta_{s-} \\
\sin \sqrt{\pi} \theta_{s+} \sin \sqrt{\pi}\phi_{s-}  \\
\cos \sqrt{\pi} \theta_{s+} \sin \sqrt{\pi}\phi_{s-} 
\end{array} \right.
\end{array} \right\},
\end{eqnarray}
where $\Gamma$ equals 
$i\eta_{\uparrow}(1) \eta_{\uparrow}(2)$ for
the singlet and $z$-component of the triplet order parameters,
and $i\eta_{\uparrow}(1) \eta_{\downarrow}(2)$ 
for $x,y$-components of the triplet order parameters,
and the same as below.

The difference of the bond strength between two legs is
$\sum_{j\sigma} (-)^{j+1} c^\dagger_{j\sigma}(i) c_{j\sigma}(i+1) +h.c.
= \sum_\sigma \psi^\dagger_{1\sigma}(x) \psi_{2\sigma}(x+a)
+ \psi^\dagger_{2\sigma}(x) \psi_{1\sigma}(x+a) +h.c. $,
similar is its triplet analog.
Their  staggered parts, $O_{SP}$ and $\vec{O}^t_{SP}$ 
\begin{eqnarray}
\left. \begin{array}{r}
O_{SP}(x)\\
\vec{O}^t_{SP,z,x,y}(x)\\
\end{array}\right\}
&=&
(-)^x 2 \sin(k_{f1}+\frac{\pi}{ 2} \delta)
i \{ e^{-i \pi \delta x-i\delta \pi/2 }
\left \{ \begin{array}{c}
\psi^\dagger_{1L\sigma}(x)\psi_{2R\sigma}(x)+
\psi^\dagger_{2L\sigma}(x)\psi_{1R\sigma}(x)-h.c.\\
\psi^\dagger_{1L\alpha}(x)( {\vec\sigma/ 2})_{\alpha\beta}
\psi_{2R\beta}(x)+
\psi^\dagger_{2L\alpha}(x) ( {\vec\sigma/ 2})_{\alpha\beta}
\psi_{1R\beta}(x)-h.c.\nonumber\\
\end{array}\right. \nonumber\\
&\propto& \frac{ 2 \Gamma }{ \pi a}
\big \{ \cos (\sqrt{\pi} \phi_{c+} -\delta \pi x-\delta \pi/2   )
\left\{\begin{array}{r}
2 \sin \sqrt{\pi} \theta_{c-}
\sin \sqrt{\pi} \phi_{s+} \cos \sqrt{\pi}\theta_{s-} \\
\cos \sqrt{\pi} \theta_{c-}
\left\{\begin{array}{r}
\sin \sqrt{\pi}  \phi_{s+}   \sin \sqrt{\pi}\theta_{s-}\\
\sin \sqrt{\pi}  \theta_{s+} \sin \sqrt{\pi}\phi_{s-}  \\
\cos \sqrt{\pi}  \theta_{s+} \sin \sqrt{\pi}\phi_{s-} \\
\end{array}\right. \\
\end{array}\right. \nonumber\\
&&  +
 \sin( \sqrt{\pi}  \phi_{c+}-\delta \pi x-\delta \pi/2  ) 
\left \{ \begin{array}{r}
2 \cos \sqrt{\pi}\theta_{c-}
  \cos \sqrt{\pi}\phi_{s+} \sin \sqrt{\pi} \theta_{s-} \\
 \sin \sqrt{\pi} \theta{c-}
\left \{ \begin{array}{r}
\cos \sqrt{\pi} \phi_{s+}    \cos \sqrt{\pi} \theta_{s-} \\
\cos \sqrt{\pi} \theta_{s+}  \cos \sqrt{\pi}\phi_{s-}  \\
-\sin \sqrt{\pi} \theta_{s+} \cos \sqrt{\pi}\phi_{s-} 
\end{array} \right.
\end{array} \right\}
\end{eqnarray}
It is clear that $O_{CDW}$ and $O_{SP}$ are real and imaginary part
of $\psi^\dagger_{1L\sigma}\psi_{2R\sigma}+
\psi^\dagger_{2L\sigma}\psi_{1R\sigma}$ respectively.

Next we present the staggered part of the diagonal current density,
$i \sum_{j} (-)^{j+1} c^\dagger_j(i) c_{j+1}(i+1)-h.c.$,
and its triplet analog as below
\bea
\left.\begin{array}{c}
O_{DC}(x)\\
\vec{O}^t_{DC,z,x,y}(x)
\end{array} \right \}
&=& (-)^x \sin(k_{f1}+\delta \pi/2)
\big \{ e^{-i\pi\delta x-i\delta/2 \pi}
\left \{ \begin{array}{l}
\psi^\dagger_{1L\sigma}(x)\psi_{2R\sigma}(x)-
\psi^\dagger_{2L\sigma}(x) \psi_{1R\sigma}(x)) +h.c.\\
\psi^\dagger_{1L\alpha}(x)({\vec\sigma/ 2})_{\alpha\beta}
\psi_{2R\beta}(x)-
\psi^\dagger_{2L\alpha}(x)({\vec\sigma/ 2})_{\alpha\beta}
\psi_{1R\beta}(x)) +h.c.\\
\end{array}\right. \nonumber \\
& \propto& \frac{2\Gamma }{ \pi a} 
\big \{ \cos (\sqrt{\pi} \phi_{c+}-\pi \delta x-\frac{\delta \pi}{ 2}) 
\left\{ \begin{array}{l}
2 \cos \sqrt {\pi} \theta_{c-}
\sin \sqrt {\pi} \phi_{s+} \cos \sqrt {\pi} \theta_{s-}  \\
-\sin \sqrt {\pi} \theta_{c-}
\left \{ \begin{array} {r} 
\sin \sqrt {\pi} \phi_{s+} \sin \sqrt {\pi} \theta_{s-}  \\
\sin \sqrt {\pi} \theta_{s+} \sin \sqrt {\pi} \phi_{s-}  \\
\cos \sqrt {\pi} \theta_{s+} \sin \sqrt {\pi} \phi_{s-}   \\
\end{array} \right.
\end{array} \right.\nonumber \\
&+&
 \sin (\sqrt{\pi} \phi_{c+}-\pi \delta x-\frac{\delta \pi}{ 2}) 
\left\{ \begin{array}{l}
-2 \sin \sqrt {\pi} \theta_{c-}
\cos \sqrt {\pi} \phi_{s+} \sin \sqrt {\pi} \theta_{s-}      \\
\cos \sqrt {\pi} \theta_{c-}
\left \{ \begin{array} {r} 
 \cos \sqrt {\pi} \phi_{s+}   \cos \sqrt {\pi} \theta_{s-}   \\
 \cos \sqrt {\pi} \theta_{s+} \cos \sqrt {\pi} \phi_{s-}   \\
-\sin \sqrt {\pi} \theta_{s+}   \cos \sqrt {\pi} \phi_{s-}   \\
\end{array} \right.
\end{array} \right\}
\end{eqnarray}

The difference of the current density along the legs is $ i \sum_j (-)^{j+1}
 ( c^\dagger_{j\sigma} (i) c_{j\sigma}(i+1) -h.c.)$.
Its staggered part is
\begin{eqnarray}
&(-)^x & 2 \cos(k_{1f} +\frac{\delta \pi }{ 2})
 (-i) \sum_\sigma \big \{ e^{-i \pi\delta x -i\delta \pi / 2 }  
(\psi^\dagger_{1\sigma L}(x)\psi_{2\sigma R}(x)
-\psi^\dagger_{2\sigma L}(x)\psi_{1\sigma R}(x)) -h.c. \big \}\nonumber. 
\end{eqnarray}

Similarly, the staggered  current along the rung 
$i (c^\dagger_{2\sigma} (i) c_{1\sigma}(i) -h.c.)$ is:
\begin{eqnarray}
(-)^x t_\perp \{ \frac{i }{ 2}\sum_{\delta} e^{-i \pi\delta x }   
(\psi^\dagger_{1\sigma L}\psi_{2\sigma R}
-\psi^\dagger_{2\sigma L}\psi_{1\sigma R} ) -h.c. \big \}\nonumber
\end{eqnarray}
It can be shown that they satisfy the continuous relation \cite{fjarestad},
so does its triplet counterpart, $\vec{O}_{DDW^t}$
staggered currents along legs and rungs have
the d-wave feature. 
We use  currents along the rung as order parameters.
Their bosonized forms are
\begin{eqnarray}
\left.\begin{array}{r}
O_{DDW}(x)\\
\vec{O}_{DDW^t,z,x,y}(x)
\end{array}
\right \} 
&\propto&
\frac{2 \Gamma }{ \pi a}
\big \{ \cos (\sqrt{\pi} \phi_{c+}-\pi \delta x) 
\left \{ \begin{array}{r}
2 \sin \sqrt {\pi} \theta_{c-} 
\cos \sqrt {\pi} \phi_{s+} \sin \sqrt {\pi} \theta_{s-}  \\
-\cos \sqrt {\pi} \theta_{c-}
\left \{ \begin{array}{r}
 \cos \sqrt {\pi} \phi_{s+}   \cos \sqrt {\pi} \theta_{s-} \\
 \cos \sqrt {\pi} \theta_{s+} \cos \sqrt {\pi} \phi_{s-} \\
-\sin \sqrt {\pi} \theta_{s+} \cos \sqrt {\pi} \phi_{s-}\\
\end{array} \right.
\end{array} \right. \nonumber \\
&+&\sin (\sqrt{\pi} \phi_{c+}-\pi \delta x)  
\left \{ \begin{array}{r} 
2  \cos \sqrt {\pi} \theta_{c-} 
\sin \sqrt {\pi} \phi_{s+}  \cos \sqrt {\pi} \theta_{s-}  \\
- \sin \sqrt {\pi} \theta_{c-}
\left \{ \begin{array}{r} 
\sin \sqrt {\pi} \phi_{s+}   \sin \sqrt {\pi} \theta_{s-} \\
\sin \sqrt {\pi} \theta_{s+} \sin \sqrt {\pi} \phi_{s-} \\
\cos \sqrt {\pi} \theta_{s+} \sin \sqrt {\pi} \phi_{s-}\\
\end{array} \right.
\end{array} \right\}.
\end{eqnarray}
It can also be seen that 
the DC and DDW order parameters are the real and imaginary parts
of $\psi^\dagger_{1L\sigma}\psi_{2R\sigma}-
\psi^\dagger_{2L\sigma}\psi_{1R\sigma}$, respectively.

Finally, the bosonized forms of the $d$-wave and $s$-wave pairing order parameters are
\begin{eqnarray}
\Delta_d&=&(\psi_{1L\uparrow}\psi_{1R\downarrow}-
\psi_{1L\downarrow}\psi_{1R\uparrow} )-
(\psi_{2L\uparrow}\psi_{2R\downarrow}-
\psi_{2L\downarrow}\psi_{2R\uparrow} )\nonumber\\
&=& \frac{2\eta_\uparrow(1)\eta_\downarrow(1) }{\pi a}
e^{i\sqrt{\pi}\theta_{c+} }
(-\cos \sqrt{\pi} \theta_{c-} \sin \sqrt{\pi} \phi_{s+}
\sin \sqrt{\pi} \phi_{s-}+i \sin \sqrt{\pi} \theta_{c-} \cos \sqrt{\pi} 
\phi_{s+} \cos \sqrt{\pi} \phi_{s-}\big )\nonumber \\
\Delta_s&=&(\psi_{1L\uparrow}\psi_{1R\downarrow}-
\psi_{1L\downarrow}\psi_{1R\uparrow} )+
(\psi_{2L\uparrow}\psi_{2R\downarrow}-
\psi_{2L\downarrow}\psi_{2R\uparrow} )\nonumber\\
&=&\frac{2\eta_\uparrow(1)\eta_\downarrow(1) }{\pi a}
e^{i\sqrt{\pi}\theta_{c+} } 
\big (\cos \sqrt{\pi} \theta_{c-} \cos \sqrt{\pi} \phi_{s+}
\cos \sqrt{\pi} \phi_{s-}
+i \sin \sqrt{\pi} \theta_{c-} \sin \sqrt{\pi} \phi_{s+}
\sin \sqrt{\pi} \phi_{s-}
\big )
\end{eqnarray}

\end{widetext}

\end{document}